\documentclass[twocolumn]{aastex62}

\usepackage{txfonts}
\usepackage{color,bm}

\newcommand{\eqref}[1]{(\ref{#1})}

\shorttitle{Atmospheric dynamics on eccentric-tilted exoplanets}
\shortauthors{Ohno \& Zhang}

\begin{document}

\title{Atmospheres on Nonsynchronized Eccentric-tilted Exoplanets I: Dynamical Regimes}

\author{Kazumasa Ohno}
\affil{Department of Earth and Planetary Sciences, Tokyo Institute of Technology, Meguro, Tokyo, 152-8551, Japan}

\author{Xi Zhang}
\affil{Department of Earth and Planetary Sciences, University of California Santa Cruz, 1156 High St, Santa Cruz, CA 95064, USA }

\begin{abstract}
Relatively long-period nonsynchronized planets---such as warm Jupiters---potentially retain the primordial rotation, eccentricity, and obliquity that might encapsulate information on planetary climate and formation processes.
To date, there has not been a systematic study on climate patterns on these planets that will significantly influence their observations.
Here we investigate the atmospheric dynamics of nonsynchronized, fast-rotating exoplanets across various radiative timescales, eccentricities, and obliquities using a shallow water model. 
The dynamical pattern can be demarcated into five regimes in terms of radiative timescale $\tau_{\rm rad}$ and obliquity $\theta$. 
An atmosphere with $\tau_{\rm rad}$ shorter than a planetary day usually exhibits a strong day--night temperature contrast and a day-to-night flow pattern. 
In the intermediate $\tau_{\rm rad}$ regime between a planetary day and a year, the atmosphere is dominated by steady temperature and eastward jet patterns for $\theta\leq{18}^{\circ}$ but shows a strong seasonal variation for $\theta\geq{18}^{\circ}$ because the polar region undergoes an intense heating at around the summer solstice. 
If $\tau_{\rm rad}$ is larger than a year, seasonal variation is very weak. 
In this regime, eastward jets are developed for $\theta\leq{54}^{\circ}$ and westward jets are developed for $\theta\geq{54}^{\circ}$. 
These dynamical regimes are also applicable to the planets in eccentric orbits.
The large effects of exoplanetary obliquities on circulation patterns might offer observational signatures, which will be investigated in Paper II of this study.
\end{abstract}

\section{Introduction} \label{sec:intro}
Since the first detection of an exoplanet \citep{Mayor&Queloz95}, observational efforts have greatly extended the known exoplanet catalog to date, which enables us to focus on not only close-in exoplanets but also outer exoplanets.
The close-in exoplanets are predicted to be in a synchronous state of their rotation by the stellar tides \citep{Guillot96,Rasio+96}.
By contrast, as orbital distance increases, planets tend to retain their primordial rotation, eccentricity, and obliquity, which is defined as the angle between the orbital normal and the spin axis of a planet.
For example, solar system planets exhibit a variety of obliquities, such as ${\sim}0^{\circ}$ (Mercury and Jupiter), ${20}^{\circ}$--${30}^{\circ}$ (Earth, Mars, Saturn, and Neptune), ${\sim}{90}^{\circ}$ (Uranus), and ${\sim}{180}^{\circ}$ (Venus).
For eccentricity, exoplanets orbiting beyond $0.03~{\rm AU}$ are observed to have various eccentricities ranging from $0$ to $1.0$ \citep{Winn&Fabrycky15}.
As an extreme example, the hot Jupiter HD 80606 b has an eccentricity of $0.93$ \citep{Moutou+09}.

Planetary obliquity is of a great interest in a number of astrophysical problems.
The planet obliquity potentially provides clues to climate and formation processes of exoplanets.
Obliquity affects the spatial distribution of the incoming stellar insolation, and hence the climate and habitability of exoplanets \citep[e.g.,][]{Williams&Kasting97, Williams&Pollard03,Kane&Torres17}.
If a high planetary obliquity is maintained by a secular spin-orbit resonance with an outer planet, it has significant effects on the tidal heating and orbital evolution of close-in exoplanets \citep{Winn&Holman05,Millholland&Laughlin18}.
Planetary obliquities have also been used to infer the dynamical histories of giant planets in the solar system \citep{Brasser&Lee15}.
If a planet has a substantially large obliquity, it might imply past giant impact events \citep[e.g.,][]{Chambers01,Kokubo&Ida07} as suggested for the origin of Uranian high obliquity \citep{Slattery+92,Kegerreis+18,Kurosaki&Inutsuka19}.

On the other hand, observables of exoplanets such as transit light curves are greatly modified by atmospheric dynamics, which are significantly influenced by eccentricity and obliquity.
A number of previous studies have thoroughly investigated atmospheric dynamics on close-in exoplanets with zero obliquity \citep[e.g.,][]{ShowmanGuillot02,Cooper&Showman05,Dobbs-Dixon&Lin08,Showman+09,Heng+11,Rauscher&Menou12,Rauscher&Menou13,Kataria+14,Kataria+16,Charnay+15,KomacekShowman16,Komacek+17,Mayne+17,Zhang&Showman16,Zhang&Showman18}.
Recent expansion of the exoplanet catalog has also motivated investigations of atmospheric dynamics on nonsynchronized exoplanets.
\citet{Showman+15} examined the influences of planetary rotation on atmospheric dynamics on warm and hot Jupiters \citep[see also][for Earth-like exoplanets]{Penn&Vallis17,Penn&Vallis18}.
They showed that the dynamics of a nonsynchronized planet is dominated by either an equatorial superrotating jet or midlatitude jets, depending on the rotation period and incident flux.
\citet{Kataria+13} investigated the atmospheric dynamics and thermal light curves of eccentric hot Jupiters \citep[see also][]{Langton&Laughlin08,Lewis+10,Lewis+14,Lewis+17}.
They found that the shape of the light curve is significantly influenced by the eccentric orbit, although the circulation patterns are qualitatively similar to that of planets in circular orbits.
But all of the above studies have assumed zero planetary obliquity.

Pioneering study of \citet{Langton&Laughlin07} investigated the atmospheric circulations on hot Jupiter with obliquity of $90$ deg using a shallow water model. 
They showed that the temperature patterns on highly tilted planets are periodic and more symmetric than that on planets with zero obliquities.
Recently, \citet{Rauscher17} investigated the dynamics on planets for a variety of obliquities and showed that the atmospheric flow pattern significantly varies with obliquity.
It was also suggested that the seasonal variation occurs when the planetary obliquity is higher than ${\sim}{30}^{\circ}$.
However, \citet{Rauscher17} assumed that the atmospheric dynamic is controlled by diurnally averaged insolation, which might not be true for planets with different insolation.
Moreover, \citet{Rauscher17} only investigated the dynamics on a planet with a circular orbit.
Because orbital eccentricity is much more difficult to damp than planetary obliquity by the stellar tides during planetary migration \citep{Peale99}, it is expected that tilted \footnote{In this study, ''tilted'' does not mean the inclined orbital plane, namely a nonzero orbital inclination. Here "tilted" means that the planet rotation axis is misaligned to its orbital normal.} planets are also likely to have nonzero eccentricities.
To date, there has not been a systematic investigation on atmospheric regimes with both nonzero eccentricity and obliquity.

We aim to investigate the atmospheric dynamics on generic eccentric-tilted exoplanets (ET planets hereafter) and its observational implications. 
Our study will be presented in two consecutive papers. 
In Paper I (the current paper), we will investigate how the dynamical regimes vary with planetary obliquity and eccentricity.
In Paper II \citep{Ohno&Zhang19}, we will calculate synthetic thermal light curves of ET planets and discuss how to potentially infer the obliquity from observations.
The organization of this paper is as follows.
We present the theoretical arguments on the dynamical regimes of ET planets in Section \ref{sec:Regime}.
We introduce our numerical model description and procedures in Section \ref{sec:method}. 
We show the dynamical patterns and thermal structures of ET planets in Section \ref{sec:result1}. 
We summarize this paper in Section \ref{sec:summary}. 

\section{Dynamical Regimes}\label{sec:Regime}

\begin{figure*}[t]
\centering
\includegraphics[clip, width=\hsize]{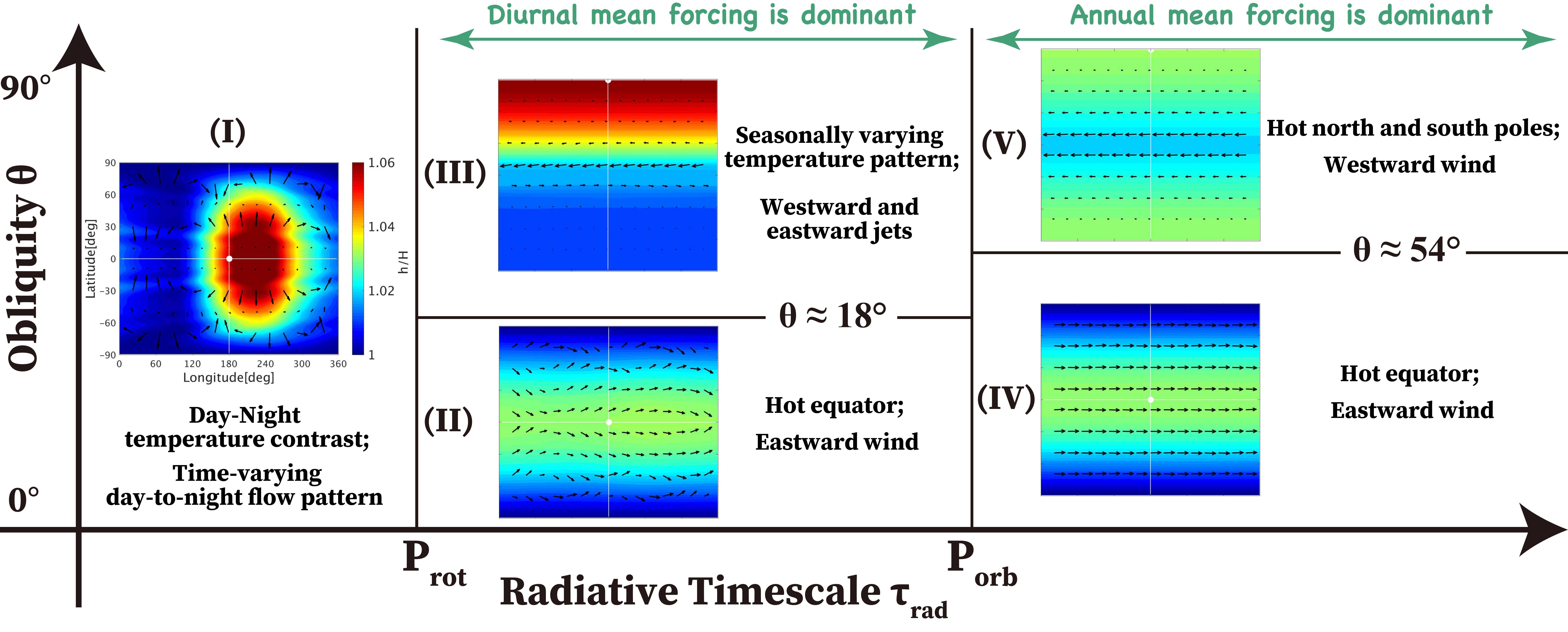}
\caption{Schematic diagrams of dynamical regimes for ET planets. The dynamical regime might be classified into five regimes in terms of the temperature and wind patterns (see Section \ref{sec:Regime}). For retrograde-rotating planets with $\theta>{90}^{\circ}$, one can translate the vertical axis to ${180}^{\circ}-\theta$ (see Section \ref{sec:Regime}). The snapshots of relevant circulation patterns taken from Section \ref{sec:result1} are also shown for each regime. }
\label{fig:regime}
\end{figure*}

\begin{figure}[t]
\centering
\includegraphics[clip, width=\hsize]{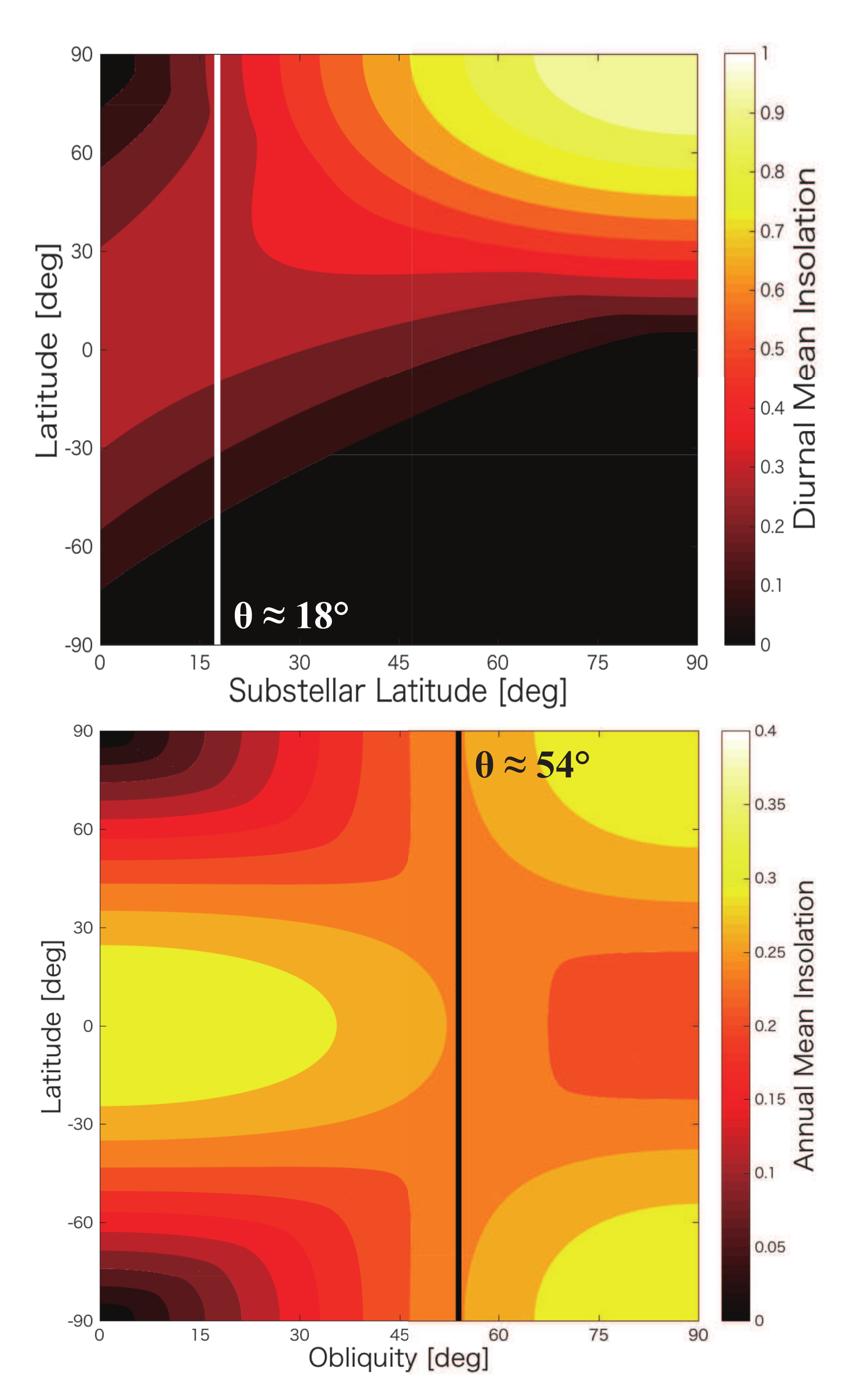}
\caption{Diurnal (top panel) and annual (bottom panel) mean insolation. 
The horizontal axis is the substellar latitude and the obliquity for the top and the bottom panels, respectively.
The vertical axes are latitude.
The color scale shows the diurnal mean insolation (Equation \ref{eq:diurnal}) normalized by $S_{\rm 0}$ and the annual mean insolation (Equation \ref{eq:annual}) normalized by $L/4\pi a^2$ in the top and the bottom panels, respectively. The white and black lines denote the critical obliquity for the transition of the dynamical regimes (see Section \ref{sec:Regime}).}
\label{fig:insolation}
\end{figure}

In this study, we classify the atmospheric dynamics of ET planets into five typical regimes in terms of planetary obliquity $\theta$ and radiative timescale $\tau_{\rm rad}$ (see Figure \ref{fig:regime}).
Since we focus on planets with a relatively long orbital period, we assume that the planetary rotation period $P_{\rm rot}$ is much shorter than the orbital period $P_{\rm orb}$.
This is true for solar system planets except for Venus, where atmospheric tides might be important \citep[e.g.,][]{Goldreich&Peale70} but are not included here.
The radiative timescale usually increases with increasing orbital distance as the planet gets colder but also depends on a number of parameters, such as vertical distributions of temperature and opacity in the atmosphere.
Because of the uncertainty of the radiative timescale, we vary $\tau_{\rm rad}$ as a free parameter in this study.

The dynamical regimes can be classified using the temperature patterns and dominant stellar heating patterns.
If $\tau_{\rm rad}$ is shorter than $P_{\rm rot}$, we expect that the temperature pattern has a strong day--night contrast.
If $\tau_{\rm rad}$ is longer than $P_{\rm rot}$ but shorter than $P_{\rm orb}$, the temperature patterns are controlled by diurnal mean insolation patterns.
If $\tau_{\rm rad}$ is longer than $P_{\rm orb}$, the temperature patterns are eventually dominated by annual mean insolation patterns.
In the latter two cases, we further expect different behaviors between the low-obliquity and high-obliquity cases.
For fast-rotating planets, if the frictional drag in the atmosphere is weak, we crudely predict that the flow patterns are largely controlled by the balance between the Coriolis force and the pressure gradient, called the geostrophic balance \citep{Vallis06}.
Here we first summarize the five dynamical regimes for tilted planets (Figure \ref{fig:regime}).
The criteria $\theta={18}^{\circ}$ and ${54}^{\circ}$ will be derived and discussed later.
\begin{itemize}
\item[(I)] When the stellar irradiation is very strong, the radiative timescale becomes short, and an instantaneous insolation pattern (a day--night heating pattern) controls atmospheric circulation.
The temperature pattern shows a hot dayside and a cold nightside. 
The wind pattern is generally dominated by substellar-to-antistellar flows. 
Both the temperature and flow patterns are highly time-dependent, slaved to the substellar point movement.
The criterion of this regime is given by $\tau_{\rm rad}< P_{\rm rot}$.
\item[(II)] 
When stellar irradiation is weak or the planetary rotation is rapid, the atmosphere does not reach the radiative equilibrium during the planetary day, and thus diurnally averaged insolation dominates the heating pattern.
If the obliquity is low, the temperature pattern shows a hot equator region and cold poles, and the seasonal variation is weak. The wind pattern is dominated by eastward flows. The criteria of regime II are $P_{\rm rot}<\tau_{\rm rad}<P_{\rm orb}$ and $\theta<{18}^{\circ}$.
\item[(III)]  Planets with large obliquities can receive more insolation at the pole than the equator during the planetary day. As a result, the temperature pattern exhibits significant seasonal variations and shows a hot illuminated summer pole and a cold equator when the planet is around the solstices. In contrast to regime II, westward flows emerge on the illuminated hemisphere in this regime because the pressure gradient is from the illuminated pole to the equator. The criteria of regime III are $P_{\rm rot}<\tau_{\rm rad}<P_{\rm orb}$ and $\theta>{18}^{\circ}$.
\item[(IV)] When the stellar irradiation is very weak, the radiative timescale is significantly long, and the atmosphere does not reach the radiative equilibrium even in a planetary year. In this situation, the stellar heating pattern is dominated by annual mean insolation rather than the diurnal mean. 
The temperature and flow patterns are similar to those in regime II with small obliquities; however, the criterion is different. Under annual mean insolation, the criterion of regime IV is given by $\tau_{\rm rad}>P_{\rm orb}$ and $\theta<{54}^{\circ}$.
\item[(V)] Planets with large obliquities receive more insolation at both poles than at the equator when averaged in a planetary year. In other words, the annual mean insolation is maximized at both poles if the obliquity exceeds a threshold. This forcing pattern induces a pressure gradient from the poles to the equator, and thus the westward flows emerge on the entire planet. The criteria of regime V is given by $\tau_{\rm rad}>P_{\rm orb}$ and $\theta>{54}^{\circ}$.
\end{itemize} 
For planets in an eccentric orbit, the eccentricity has relatively minor effects because it only affects the magnitude of temperature but not the shape of the temperature distribution.
However, the eccentricity might lead to a transition from regime III to V as discussed in Section \ref{sec:regime23_e05}. 
Here, we explain the physical basis of each regime in detail.

In regime I, the dynamical pattern is controlled by the day--night heating.
A specific case of this regime---the zero-obliquity case---corresponds to the highly illuminated regime examined by \citet{Showman+15}.
In the case of nonzero obliquity, the temperature patterns should be significantly different from that of nontilted planets. 
For a high-obliquity planet, the substellar point moves across a large latitudinal range with time.
Because the Coriolis force also changes substantially with latitude, we expect that the flow pattern behaves very differently in different seasons.

In regimes II and III, we predict that the dynamical patterns should be controlled by diurnal mean insolation. 
The parameter space investigated by \citet{Rauscher17} is located in these two regimes.
The diurnal mean insolation $S_{\rm d}(\phi)$ as a function of latitude is expressed as \citep[e.g.,][]{Williams&Kasting97}
\begin{equation}\label{eq:diurnal}
S_{\rm d}(\phi)=\frac{S_{\rm i}}{\pi}(D_{\rm h}\sin{\phi}\sin{\phi_{\rm ss}}+\cos{\phi}\cos{\phi_{\rm ss}}\sin{D_{\rm h}}),
\end{equation} 
where $S_{\rm i}$ is the incoming stellar flux, $\phi$ is the latitude, $\phi_{\rm ss}$ is the substellar latitude, and $D_{\rm h}$ is the hour angle between sunset and sunrise, given by
\begin{equation}
\cos{D_{\rm h}}=-\tan{\phi}\tan{\phi_{\rm ss}},~{\rm for~}0<D_{\rm h}<\pi,
\end{equation}
where $\phi_{\rm ss}=\pm \theta$ stands for the solstices and $\phi_{\rm ss}=0$ for the equinox.
The eccentricity only affects the incoming flux $S_{\rm i}$ so that it varies with the star--planet distance.
We plot Equation \eqref{eq:diurnal} in the top panel of Figure \ref{fig:insolation} as a function of $\phi$ and $\phi_{\rm ss}$.
The diurnal mean insolation is maximized either at the equator ($\phi=0$) or at the illuminated pole ($\phi=+\pi/2$ or $-\pi/2$), resulting in distinct temperature patterns.
Equating $S_{\rm d}(0)$ and $S_{\rm d}(\pi/2)$, we can obtain the critical substellar latitude $\phi_{\rm d}$ for the transition between the above two types of insolation distributions:
\begin{equation}
\phi_{\rm d}={\rm tan}^{-1}\left(\frac{1}{\pi}\right)\approx {18}^{\circ}.
\end{equation}
Because the substellar latitude $\phi_{\rm d}$ can only vary from zero to planetary obliquity $\theta$, the insolation distributions are expected to be different between planets with $\theta<\phi_{\rm d}$ and $\theta>\phi_{\rm d}$.
Thus, we predict that the dynamical regime changes at around $\theta={18}^{\circ}$.
For $\theta<{18}^{\circ}$ (regime II), the insolation is maximized at the equator throughout the planet orbit, and there is no strong seasonal variation.
For $\theta>{18}^{\circ}$ (regime III), the insolation is maximized at the equator around equinoxes but at the illuminated pole around solstices.
Therefore, we expect a strong seasonal variation in circulation and temperature patterns for $\theta>{18}^{\circ}$.
\citet{Rauscher17} found that the seasonal variation is remarkable for $\theta \geq {30}^{\rm \circ}$ from 3D numerical simulations, qualitatively consistent with the obliquity criteria derived here.

The temperature patterns are controlled by annually averaged insolation rather than diurnal average when the radiative timescale is significantly longer than the orbital period in regimes IV and V.
The annual mean insolation $S_{\rm a}(\phi)$ as a function of latitude is given by \citep{Ward74}
\begin{equation}\label{eq:annual}
S_{\rm a}(\phi)={\displaystyle \frac{L_{\rm 0}}{8\pi^3a^2\sqrt{1-e^2}}\int^{2\pi}_{0}{[1-(\sin{\phi}\cos{\theta}-\cos{\phi}\sin{\theta}\sin{\psi})^2}]^{1/2}d\psi},
\end{equation}
where $L_{\rm 0}$ is the stellar luminosity and $a$ is the semi-major axis.
We plot Equation \eqref{eq:annual} in the bottom panel of Figure \ref{fig:insolation} for $e=0$.
The annual mean insolation is maximized either at the equator or {\it both} poles.
Here, $S_{\rm a}(0)$ and $S_{\rm a}(\pi/2)$ are given by
\begin{equation}\label{eq:1}
S_{\rm a}(0)=\frac{L_{\rm 0}}{2\pi^3a^2\sqrt{1-e^2}}E(\sin{\theta}),
\end{equation}
\begin{equation}\label{eq:2}
S_{\rm a}(\pi/2)=\frac{L_{\rm 0}\sin{\theta}}{4\pi^2a^2\sqrt{1-e^2}},
\end{equation}
where $E(\sin{\theta})$ is the complete elliptical integral of the second kind \citep{Ward74}.
Equating Equations \eqref{eq:1} and \eqref{eq:2}, we obtain
\begin{equation}\label{eq:theta_anu}
\sin{\theta_{\rm a}}=\frac{2}{\pi}E(\sin{\theta_{\rm a}}).
\end{equation}
The solution of Equation \eqref{eq:theta_anu} corresponds to the critical obliquity $\theta_{\rm a}$ for the annual mean insolation, which is approximately $\theta_{\rm a}\approx {54}^{\circ}$ \citep{Ward74}. 
The average insolation is maximized at the equator for $\theta<{54}^{\circ}$ (regime IV) and at both north and south poles for $\theta>{54}^{\circ}$ (regime V).
As a result, the temperature and circulation patterns in regimes IV and V are very different.

It is worth noting that both obliquity criteria, $\theta={18}^{\circ}$ and ${54}^{\circ}$, are independent of the eccentricity.
For the criterion separating regimes II and III, the eccentricity only changes the magnitude $S_{\rm i}$ and does not affect the spatial pattern of the diurnal mean insolation when the substellar latitude is the same.
For the criterion dividing regimes IV and V, latitudinal dependence of the annual mean insolation is also not influenced by the eccentricity.
Therefore, those obliquity criteria are also applicable to both the eccentric-orbit planets and the circular-orbit planets, which will be demonstrated by numerical simulations in Section \ref{sec:dynamic_ET}.
Although the above arguments were derived for $\theta \leq {90}^{\circ}$, the atmospheric dynamics on a fast-rotating planet with retrograde rotation ($\theta>{90}^{\circ}$) generally behave similar to that with ${180}^{\circ}-\theta$.
This will also be demonstrated in Section \ref{sec:dynamic_retro}.

The above regime demarcation is mainly applicable to the warm exoplanets that are not far from the host star and still receive a considerably larger incoming stellar flux than the internal flux from their deep convective interiors.
These regimes should not be simply applied to the cold giant planets, such as the four giant planets in our solar system. 
Very crudely speaking, it seems interesting that the four giant planets in our planetary system fall into different regimes in our diagram (Figure \ref{fig:regime}). 
Using the radiative timescale from \citet{Li+18}, we can roughly classify Jupiter into regime II, Saturn and Neptune into regime III, and Uranus into regime V.
Indeed, Saturn (supposed to be in regime III) shows a strong seasonal cycle \citep{Guerlet+18}, and Uranus (supposed to be in regime V) has a broad westward jet \citep{Ingersoll90,Liu&Schneider10}. 
But a careful analysis shows that their atmospheric behaviors do not precisely follow our regime classification here. 
For example, Neptune (supposed to be in regime III) shows a strong westward jet at the equator in the troposphere \citep{Ingersoll90,Liu&Schneider10}, and Jupiter’s stratosphere (supposed to be in regime II) shows a strong four-year quasi-periodic oscillation at the equator \citep{Leovy+91,Orton+91}. 
The reason why our regime classification fails in the cold giant planet regime is that the incoming stellar flux is comparable to or less than the outgoing interior flux.
In that context, the waves from convective interiors significantly affect the circulation in the upper atmosphere \citep{Conrath+90,West+92,Lian&Showman10,Showman+18}.
On the other hand, for the warm exoplanets that we focus on here, the effects of convection from the deep atmosphere should have minor impacts on the circulation because the stellar flux is stronger than the expected interior flux by several orders of magnitude \citep{Showman+15}.

\section{Model Description}\label{sec:method}
\begin{figure}[t]
\centering
\includegraphics[clip, width=\hsize]{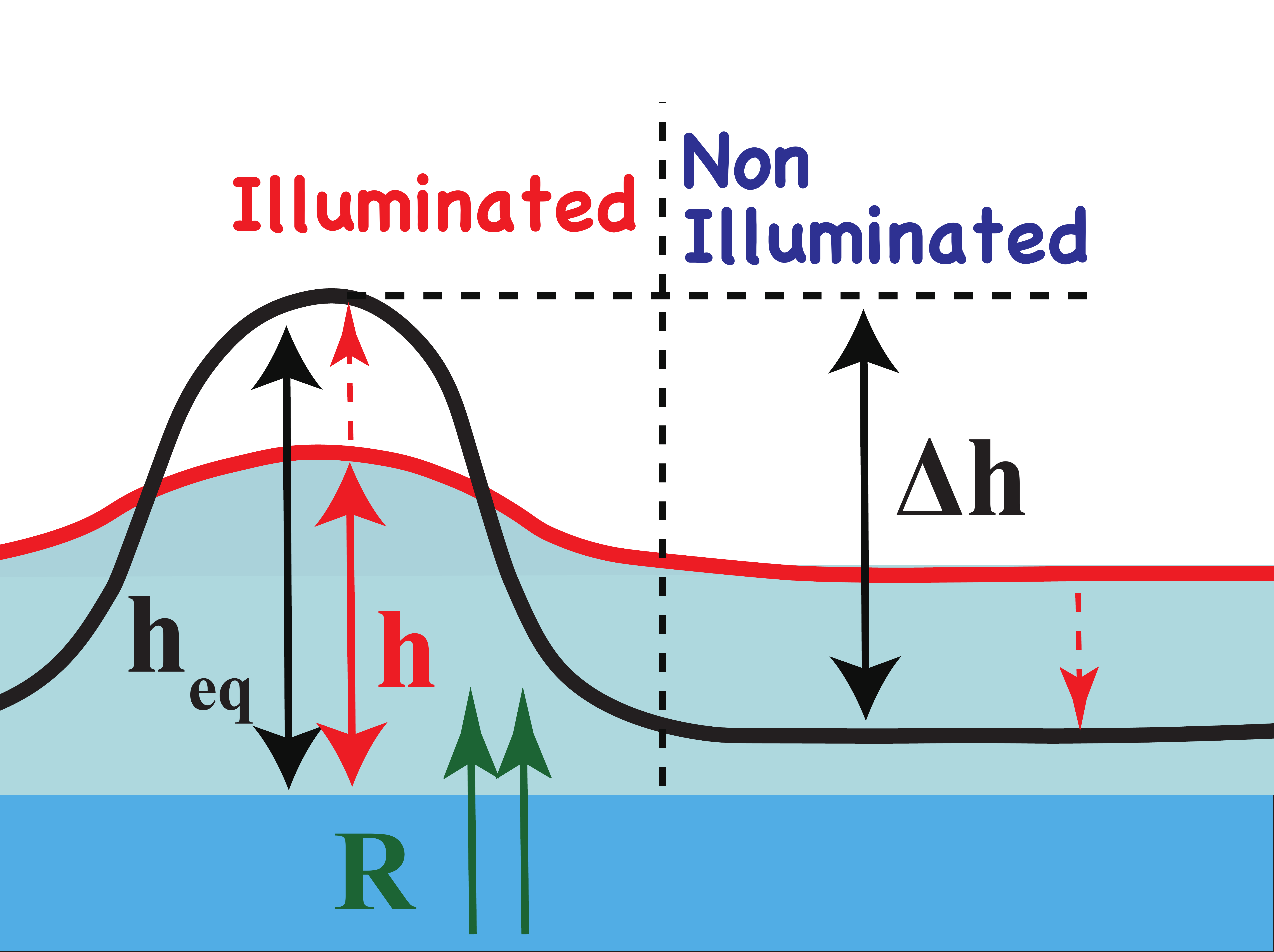}
\caption{Schematic illustration of a one-and-a-half-layer shallow water model. 
The atmosphere consists of an upper active layer (light blue layer) and a infinitely deep quiescent layer (deep blue layer). 
Net radiative heating restores the upper layer thickness $h$ (red line) to a local equilibrium value $h_{\rm eq}$ (black line), accompanied by the exchange of the mass and the momentum between the upper and deeper layers (green arrow). }
\label{fig:shallow}
\end{figure}

We use a one-and-a-half-layer shallow water model to simulate the atmospheric dynamics of nonsynchronized ET exoplanets (see Figure~\ref{fig:shallow}).
Shallow water models have been extensively used to study gas giants in the solar system \citep[e.g.,][]{Scott&Polvani08}, hot Jupiters \citep{ShowmanPolvani11,Liu&Showman13,Perez-Becker13}, brown dwarfs \citep{Zhang&Showman14}, and extraterrestrial planets \citep{Penn&Vallis17,Hammond&Pierrehumbert18}. 
Although these models are highly idealized, they are able to capture the essential features of dynamical and thermal structures.
The model also allows us to efficiently explore the atmospheric dynamics for a broad range of parameters in this study.

\begin{figure*}[t]
\centering
\includegraphics[clip, width=0.85\hsize]{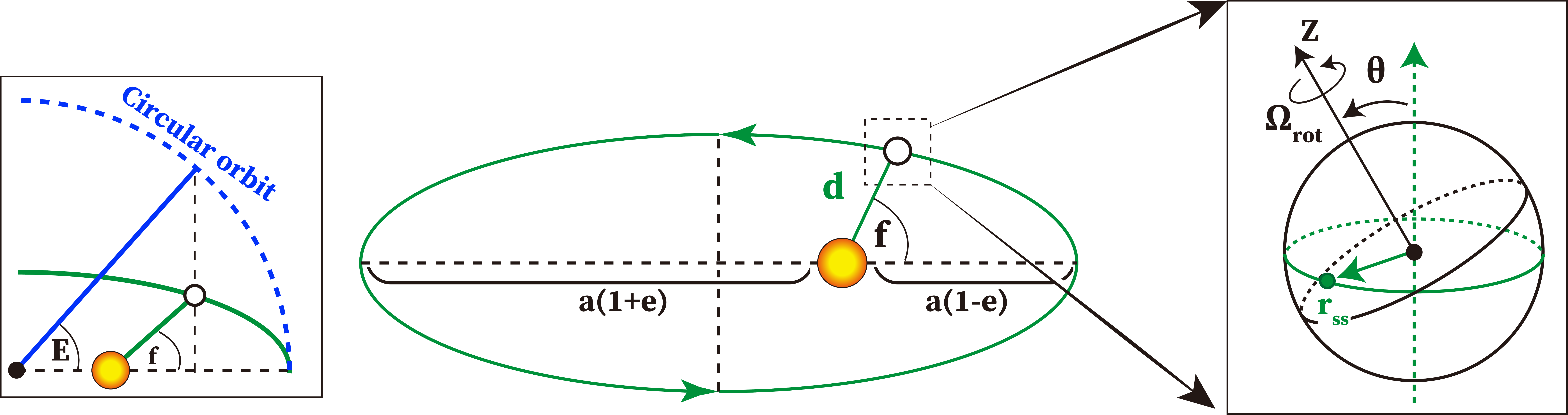}
\caption{Illustration of the eccentric-tilted planet system. The green trajectory and arrow represent the orbital plane and normal, respectively. The blue dotted curve represents the circular orbit with the same semimajor axis. Here $f$ is the true anomaly, $E$ is the eccentric anomaly, $\theta$ is the planetary obliquity, and $\mathbf{r}_{\rm ss}$ is the substellar point. }
\label{fig:illustration}
\end{figure*}

Our shallow water model is constructed of an upper active layer with a variable height $h$ and a lower quiescent layer, where each layer has a constant density.
We calculate the evolution of the height field $h(\lambda,\phi)$ and the horizontal velocity field $\mathbf{v}(\lambda,\phi)$ of the upper active layer driven by external forcing \citep[for details, see][]{Vallis06}. 
The master equations of the shallow water model, given as momentum and continuity equations, are
\begin{equation}
\label{eq:moment}
\frac{d\mathbf{v}}{dt}+g\nabla h + \mathcal{F} \mathbf{k}\times \mathbf{v}= \mathbf{R}-\frac{\mathbf{v}}{\tau_{\rm drag}},
\end{equation}
\begin{equation}
\label{eq:mass}
\frac{\partial h}{\partial t} +\nabla \cdot(\mathbf{v}h) = \frac{h_{\rm eq}(\lambda,\phi,t)-h}{\tau_{\rm rad}} \equiv Q,
\end{equation}
where $\mathbf{v}$ is the horizontal velocity vector, $\lambda$ is the longitude, $\phi$ is the latitude, $\mathbf{k}$ is the vertical unit vector, $g$ is the gravitational acceleration, $\mathcal{F}=2\Omega_{\rm rot} {\rm sin}\phi$ is the Coriolis parameter and $\Omega_{\rm rot}$ is the angular velocity of the planetary rotation, $\tau_{\rm drag}$ is the characteristic timescale of the momentum drag, and $Q$ is the net radiative heating rate of the upper layer.

The momentum exchange term $\mathbf{R}$ is designed to represent the mass transfer from the deeper quiescent layer triggered by the radiative heating. 
Following \citet{ShowmanPolvani11}, we describe $\mathbf{R}$ as
\begin{equation}
\label{eq:R}
    \mathbf{R} =  
\left\{
\begin{array}{ll}
      -Q \mathbf{v} / h  & \text{($Q>0$)} \\
      0 & \text{($Q<0$)}.
         \end{array}
\right.
\end{equation}
Note that moving out of the fluid from the upper layer does not affect the momentum of the upper layer.
This term ensures that the system reaches a single statistical equilibrium state from any initial condition for tidally locked planets \citep{Liu&Showman13}.
Also, the term is crucial for emergence of the equatorial superrotation on synchronized planets in shallow water simulations \citep{ShowmanPolvani11}.

The term of $\mathbf{v}/\tau_{\rm drag}$ is a parameterized drag term \citep[e.g.,][]{ShowmanPolvani11}. 
The drag timescale $\tau_{\rm drag}$ encapsulates various effects such as the Lorentz force drag \citep[][]{Perna+10,Rauscher&Menou13} or turbulent mixing \citep{Li&Goodman10,Youdin&Mitchell10}. 
The effect of the magnetic field on dynamics is important only at high temperatures for sufficient thermal ionization \citep[e.g.,][]{Rogers&Komacek14}, and it might have minor effects on nonsynchronized exoplanets with relatively low temperatures. 
On the other hand, subgrid turbulence caused by hydrodynamic instability could dissipate the momentum \citep[e.g.,][]{Li&Goodman10}, which may act as the drag in atmospheres on nonsynchronized exoplanets.
Therefore, we set a relatively long $\tau_{\rm drag}=1000$ days for all simulations in this study.

\begin{table*}[t]
  \caption{Model Parameters Used in This Study.}\label{table:1}
  \centering
  \begin{tabular}{c r c} \hline
    Parameter & Range & Description \\ \hline \hline
    $\theta$ & $0^{\circ}$, ${10}^{\circ}$, ${30}^{\circ}$, ${60}^{\circ}$, ${90}^{\circ}$, ${120}^{\circ}$, ${150}^{\circ}$, ${180}^{\circ}$ & Planetary obliquity\\
    $f_{\rm sol}$ & $0^{\circ}$, $\pm{45}^{\circ}$, $\pm{90}^{\circ}$, $\pm{135}^{\circ}$, ${180}^{\circ}$ & Northern summer solstice phase\\
    $e$ & $0$, $0.3$, $0.5$ & Orbital eccentricity \\
    $\tau_{\rm rad}$ & $0.1$, $5$, $100$ days & Radiative timescale \\ 
    $\tau_{\rm drag}$ & $1000$ days & Drag timescale \\ 
    $P_{\rm rot}$ & $0.5$ day & Planetary rotation period \\ 
    $P_{\rm orb}$ & $30$ days & Planetary orbital period \\
    $gH$ & $2\times {10}^{5}~{\rm m^{2}~s^{-2}}$ & Mean geopotential \\
    $\Delta h/H$ & $0.1$ & Relative forcing amplitude \\ \hline
  \end{tabular}
\end{table*} 

For radiative heating and cooling, we adopt the Newtonian cooling scheme that relaxes the height field toward a local equilibrium height distribution $h_{\rm eq}$ with a radiative timescale $\tau_{\rm rad}$ \citep[e.g.,][]{ShowmanPolvani11,Liu&Showman13,Perez-Becker13,Zhang&Showman14}. 
In this study, we parameterize the time-varying equilibrium height field $h_{\rm eq}$ as
\begin{equation}\label{eq:h_eq}
       h_{\rm eq} =  
      H+\Delta h \left( \frac{1-e^2}{1+e\cos{f}}\right)^{-2}(\mathbf{r_{\rm ss}} \cdot \mathbf{r})\mathcal{H}(\mathbf{r_{\rm ss} \cdot \mathbf{r}}),
\end{equation}
where $H$ is the mean atmospheric height on the nightside\footnote{One can also understand $H$ using $gH$, the mean geopotential on the nightside. See Equation \eqref{eq:gH} later for discussion.}, $f$ is the true anomaly; $\mathbf{r}$ is the unit point vector, $\mathbf{r}_{\rm ss}$ is the unit point vector of the substellar point, and $\Delta h$ is the difference of the equilibrium height between the substellar point and the nightside.
In this study, we set $\Delta h=0.1H$.
Here, $\mathcal{H}(x)$ is the Heaviside step function since the nightside receives no irradiation:
\begin{equation}\label{eq:Heaviside}
    \mathcal{H}(x) =  
\left\{
\begin{array}{ll}
      1  & \text{when $x \ge 0$} \\
      0 & \text{when $x < 0$}.
         \end{array}
\right.
\end{equation}
The factor of $[(1-e^2)/(1+e\cos{f})]^{-2}$ expresses the time variability of the incoming flux in an eccentric orbit.
When one calculates the true anomaly, it is convenient to introduce the eccentric anomaly $E$ \citep[][see also Figure \ref{fig:illustration}]{Murray&Dermott99}.
From a geometric argument, the eccentric anomaly $E$ is associated with the true anomaly $f$ as 
\begin{equation}\label{eq:f_E}
\tan{\frac{f}{2}}=\sqrt{\frac{1+e}{1-e}}\tan{\frac{E}{2}}.
\end{equation}
The time evolution of the eccentric anomaly can be calculated with the Kepler's equation \citep{Murray&Dermott99}
\begin{equation}\label{eq:Kepler2}
\frac{dE}{dt}=\frac{2\pi}{P_{\rm orb}}\frac{1}{1-e\cos{E}}.
\end{equation}
We calculate $f$ using Equations \eqref{eq:f_E} and \eqref{eq:Kepler2} for each time step.
To simulate the effects of obliquity, we introduce the time-varying motion of the substellar point, described as (for the derivation, see \citet{Dobrovolskis09,Dobrovolskis13})
\begin{equation}
\label{eq:sspoint}
\mathbf{r_{\rm ss}}=\left(
    \begin{array}{c}
     (1-\cos{\theta} )\cos{(f-f_{\rm sol})}\sin{\Omega_{\rm rot}t} -\sin{(\Omega_{\rm rot}t-f+f_{\rm sol})} \\
    (1-\cos{\theta} )\cos{(f-f_{\rm sol})}\cos{\Omega_{\rm rot}t} -\cos{(\Omega_{\rm rot}t-f+f_{\rm sol})} \\
    \cos{(f-f_{\rm sol})}\sin{\theta}
    \end{array}
    \right),
\end{equation}
where $f_{\rm sol}$ is the true anomaly of the northern summer solstice.
In contrast to \citet{Rauscher17}, who assumed the diurnally averaged stellar flux patterns, we explicitly calculate the diurnal cycle because some of our cases have very short radiative timescales and the assumption of a diurnal average is invalid.

For ET planets, there are two kinds of seasons: the obliquity season and the eccentricity season.
The obliquity season originates from the insolation distribution change due to the misalignment between the planetary rotation axis and the orbital normal.
The eccentricity season is from the insolation magnitude change that results from the star--planet distance change in an eccentric orbit.
The true anomaly of the northern summer solstice $f_{\rm sol}$ is used to quantify the interaction between the obliquity season and the eccentricity season.
For example, in the case of $f_{\rm sol}=0^{\circ}$, the planet experiences the northern summer solstice at the periapse and thus a very hot summer in the northern hemisphere.
In the case of $f_{\rm sol}={90}^{\circ}$, the planet experiences the vernal equinox at the periapse, which might lead to a cooler summer in the northern hemisphere than the spring at the equinox.

The mean geopotential $gH$ in our model is calculated based on the Rossby deformation radius in a 3D stratified atmosphere. 
The Rossby deformation radius is the characteristic length scale within which the buoyancy force dominates the dynamics over the Coriolis force.
Following \citet{Zhang&Showman14}, the mean geopotential is given by
\begin{equation}\label{eq:gH}
gH=\gamma \kappa^{2}c_{\rm p}T,
\end{equation}
where $c_{\rm p}$ is the specific heat of the atmosphere, $\kappa=R/\mu c_{\rm p}$ is ${\sim}2/7$ for hydrogen-dominated atmospheres, and $\gamma=[1+(c_{\rm p}/g)dT/dz]$ is the metric of subadiabaticity that equals unity for isothermal and zero for adiabatic atmospheres. 
Substituting $c_{\rm p}=1.3\times{10}^4~{\rm J~{kg}^{-1}~K^{-1}}$ for hydrogen-dominated atmospheres, $\gamma=0.1$--$1$ \citep{Zhang&Showman14}, and $T=T_{\rm eq}{\sim}650~{\rm K}$,\footnote{The equilibrium temperature is given by $T_{\rm eq}=T_{\rm *}\sqrt{R_{\rm *}/2a}$, where $T_{\rm *}$ and $R_{\rm *}$ are the stellar effective temperature and radius. Here we set $T_{\rm *}=6000~{\rm K}$ and $R_{\rm *}=7\times {10}^{8}~{\rm m}$ for sun-like stars.} we obtain $gH{\sim}7\times {10}^{4}$--${10}^{5}~{\rm m^{2}~s^{-2}}$. 
In this study, we adopt the intermediate value of $gH=2\times {10}^{5}~{\rm m^{2}~s^{-2}}$ for all simulations.

We perform simulations using the Spectral Transform Shallow Water Model \citep[STSWM;][]{STSWM} to solve the Equations (\ref{eq:moment})--(\ref{eq:mass}) in spherical coordinates. 
This model was previously used for hot Jupiters and brown dwarfs \citep{ShowmanPolvani11,Perez-Becker13,Zhang&Showman14}.
The equations are integrated with the spectral resolution of T170, which is equivalent to 512 longitude $\times$ 256 latitude. 
The numerical stability is maintained by $\nabla^6$ hyperviscosity, where we set the hyperviscosity coefficient $\nu_{\rm 6}={10}^{30}~{\rm m^6~s^{-1}}$ so that the smallest scales are diffused with a timescale of ${\sim}{10}~{\rm days}$.
Although larger hyperviscosity leads to more smoothed height and velocity fields, the large-scale structures focused on this study are hardly influenced by the choice of hyperviscosity.
We assume the rotation period $P_{\rm rot}=0.5$ day \footnote{In this study, one day refers to one Earth day, namely $86,400~{\rm s}$.} and the planetary radius $R_{\rm p}=8.2\times {10}^{7}~{\rm m}$, values similar to that of Jupiter.
It is suggested that, for Jupiter-like planets around a solar-mass star, the tidal spin-down timescale is longer than the typical system age (${\sim}{10}^{9}~{\rm yr}$) beyond $0.1$--$0.2~{\rm AU}$ \citep{Showman+15, Rauscher17}.
Therefore, we set the orbital period $P_{\rm orb}=30$ days, equivalent to the orbital period at $a=0.2~{\rm AU}$ around sun-like stars, at which the tidal synchronization might not be effective for Jupiter-like planets.
We carry out numerical simulations for possible combinations of $\tau_{\rm rad}=0.1$, $5$, $100$ days; $e=0$, $0.3$, $0.5$; $\theta=0^{\circ}$, ${10}^{\circ}$, ${30}^{\circ}$, ${60}^{\circ}$, ${90}^{\circ}$, ${120}^{\circ}$, ${150}^{\circ}$, ${180}^{\circ}$; and $f_{\rm sol}=0^{\circ}$, $\pm{45}^{\circ}$, $\pm{90}^{\circ}$, $\pm{135}^{\circ}$, ${180}^{\circ}$.
We have run the simulations over $500$ days for $\tau_{\rm rad}=0.1$ and $5~{\rm days}$ and $1500$ days for $\tau_{\rm rad}={100}~{\rm days}$, and we ensure that the system reaches a steady state.
The parameters are summarized in Table \ref{table:1}.

\section{Atmospheric Dynamical Patterns and Thermal Structures}\label{sec:result1}
\begin{figure*}[t]
\centering
\includegraphics[clip, width=\hsize]{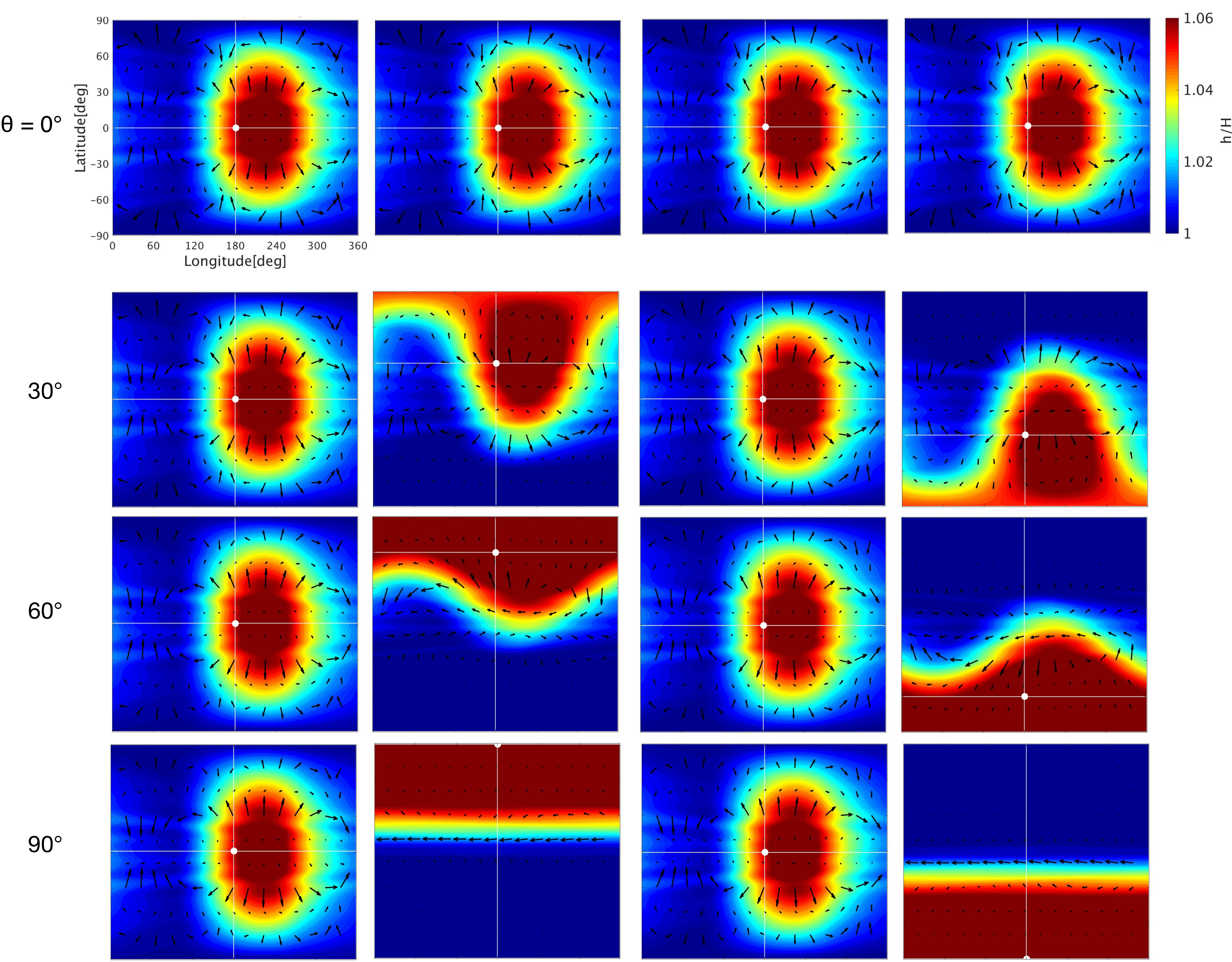}
\caption{Snapshots of height fields (color scale) and wind velocities (arrows) for planets with $e=0$. The radiative timescale is set as $\tau_{\rm rad}=0.1~{\rm day}$ in regime I, as presented in Section \ref{sec:Regime}. The height fields $h$ are normalized by the mean atmospheric height $H$ on the nightside, and thus the color scale is always larger than unity (see Equation \ref{eq:h_eq}). The horizontal axes are longitude chosen so that the substellar point, denoted as the white dots, is placed at ${180}^{\circ}$.
Each column, from left to right, shows the snapshots at vernal equinox, summer solstice, fall equinox, and winter solstice, respectively. In each row, from top to bottom, the planetary obliquity is set as $0^{\circ}$, ${30}^{\circ}$, ${60}^{\circ}$, and ${90}^{\circ}$, respectively.
Note that the lengths of arrows are normalized by the maximum length at each snapshot to clarify the dynamical patterns and do not represent magnitudes of the wind velocity. }
\label{fig:hf_e0t05}
\end{figure*}

\begin{figure*}[t]
\centering
\includegraphics[clip, width=\hsize]{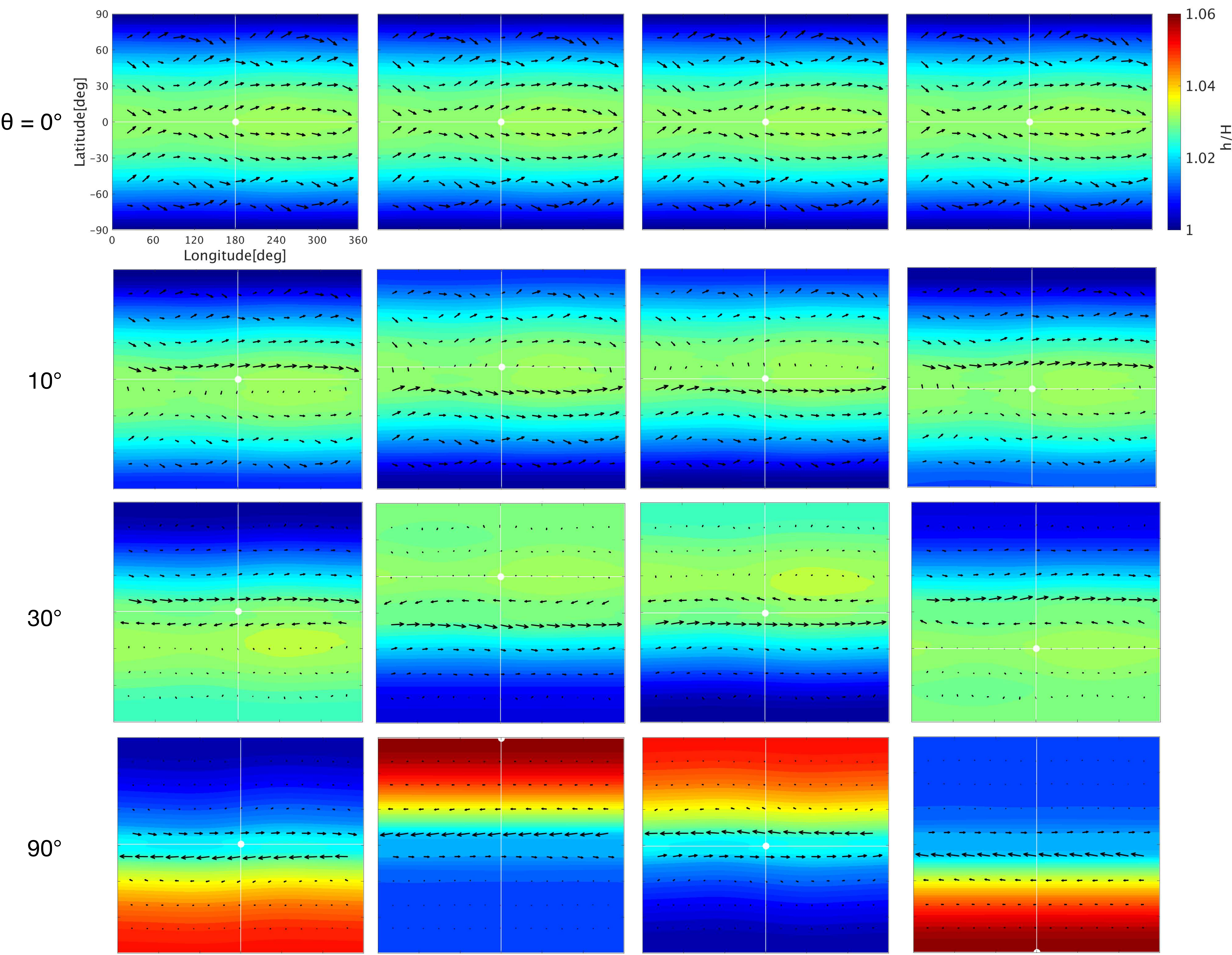}
\caption{Same as Figure \ref{fig:hf_e0t05}, but for $\tau_{\rm rad}=5~{\rm days}$ in regimes II and III in which the diurnal mean heating is dominant. From top to bottom, each row shows the snapshots for obliquity of  $0^{\circ}$, ${10}^{\circ}$, ${30}^{\circ}$, and ${90}^{\circ}$, respectively.}
\label{fig:hf_e0t5}
\end{figure*}

\begin{figure*}[t]
\centering
\includegraphics[clip, width=\hsize]{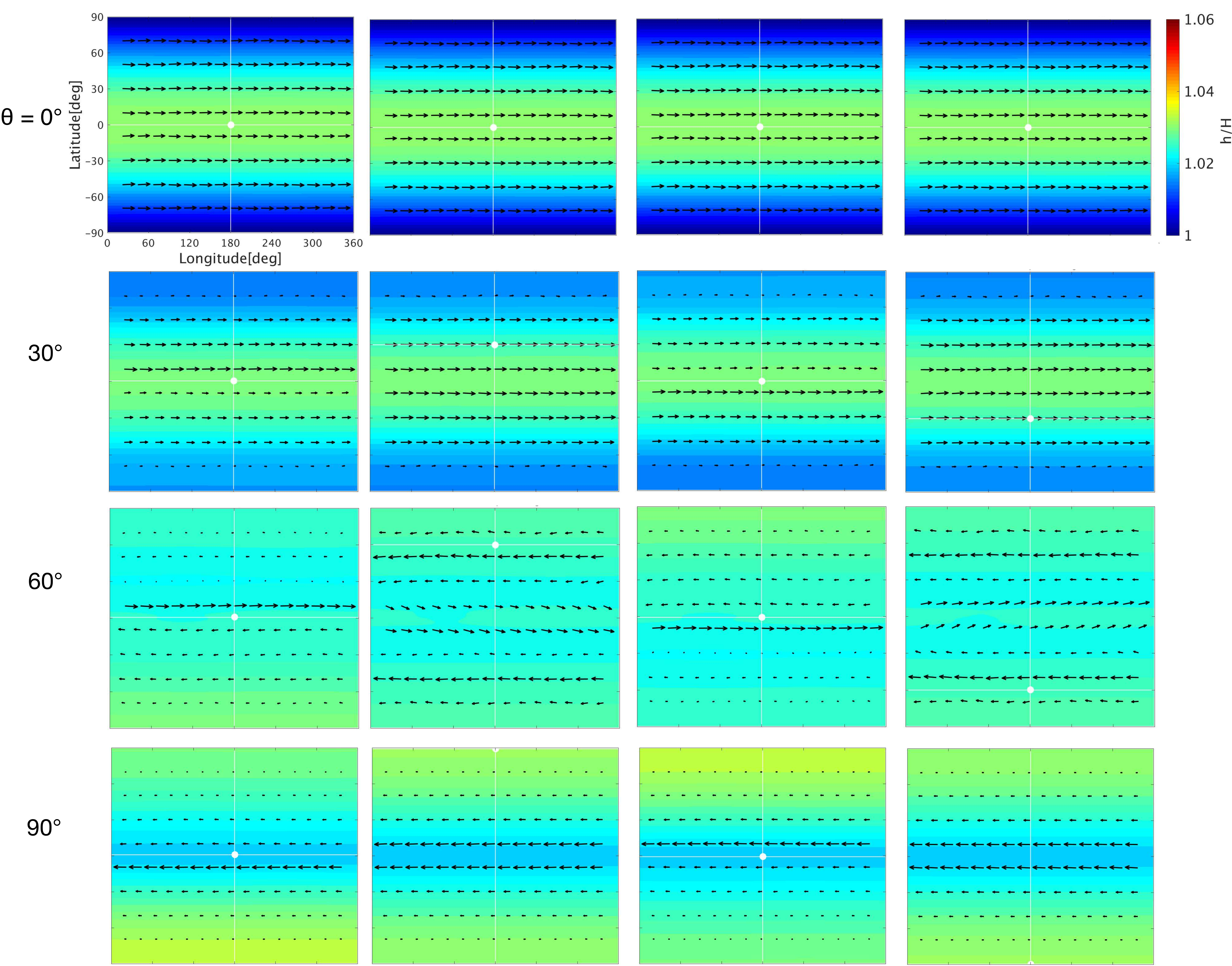}
\caption{Same as Figure \ref{fig:hf_e0t05}, but for $\tau_{\rm rad}=100~{\rm days}$ in regimes IV and V in which the annual mean heating is dominant. From top to bottom, each row shows the snapshots for obliquity of  $0^{\circ}$, ${30}^{\circ}$, ${60}^{\circ}$, and ${90}^{\circ}$, respectively. }
\label{fig:hf_e0t100}
\end{figure*}

\begin{figure*}[t]
\centering
\includegraphics[clip, width=\hsize]{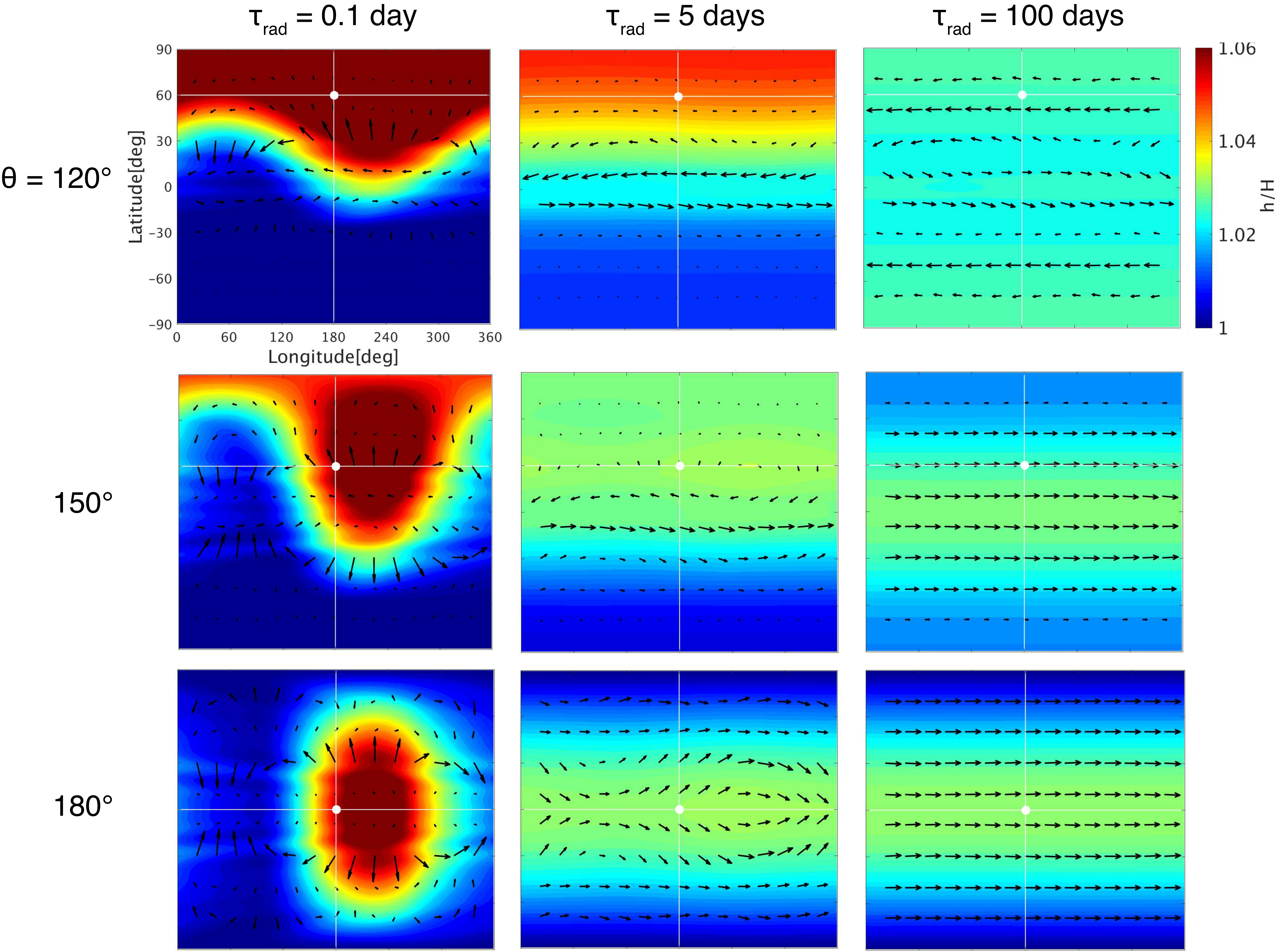}
\caption{Snapshots of height fields (color scale) and winds (arrows) for circular-orbit planets with retrograde rotation. The snapshot is taken at the northern summer solstice for each panel. Each column, from left to right, shows the fields for $\tau_{\rm rad}=0.1$, $5$, and $100~{\rm days}$, respectively. Each row, from top to bottom, shows the snapshot for $\theta={120}^{\circ}$, ${150}^{\circ}$, and ${180}^{\circ}$, respectively.}
\label{fig:retro_summary}
\end{figure*}

\subsection{Tilted Planets in Circular Orbits}\label{sec:dynamics_tilt}
We first show dynamical patterns for planets in circular orbits.
We demonstrate that height fields and flow patterns drastically vary with planetary obliquity.
Figure \ref{fig:hf_e0t05}, \ref{fig:hf_e0t5}, and \ref{fig:hf_e0t100} show the height fields and flow patterns on planets with a variety of obliquities for $\tau_{\rm rad}=0.1$, $5$, and $100~{\rm days}$, respectively.
The snapshots of height fields and flow patterns are taken at the orbital phase of the vernal equinox, northern summer solstice, fall equinox, and northern winter solstice.
Here we discuss the results in different regimes.

\subsubsection{Regime I: Day--Night Contrast Is Dominant}\label{sec:regime1_e0}
In the case of $\tau_{\rm rad}=0.1~{\rm day}$, the circulation is driven by the time-varying day--night insolation patterns (Figure \ref{fig:hf_e0t05}).
The height field exhibits a strong day--night contrast for zero obliquity (top row in Figure \ref{fig:hf_e0t05}).
This is qualitatively similar to the results of \citet{Showman+15} in which the radiative timescale is shorter than the diurnal cycle (the bottom right panel of Figure 4 in \citet{Showman+15}).
For nonzero obliquity, the hottest point is slaved by the substellar insolation that seasonally moves in both longitude and latitude.
The height field patterns at the equinox are almost the same for all obliquities because the height field is mainly determined by the equilibrium field at each moment.
By contrast, the height fields at the solstices show south--north contrast as obliquity increases.

Our simulated atmospheres mainly exhibit strong day-to-night flow patterns instead of equatorial jets.
In fact, the zonal-mean zonal winds in our simulations of nontilted planets (not shown here) show a weak jet-like structure at the equator, but the peak is much weaker compared with the 3D model results in \citet{Showman+15}, even though our height fields look similar to their temperature distributions.
Some cases with short radiative timescales in \citet[][Figure 5, the cases in the bottom row]{Showman+15} develop strong equatorial jets, although the parameters in their 3D simulations are not exactly the same as that in our 2D simulations here.
The weak equatorial jet in our simulations in regime I is probably due to the fact that the shallow water system is not able to simulate the baroclinic dynamics, which might be crucial for jet development in fast-rotating planets with a small Rossby number \citep{Penn&Vallis17}.

Our simulations always exhibit an eastward displacement of the hot spot from the substellar point.
For tidally locked planets, a number of mechanisms producing the hot spot displacement have been proposed, for example, zonal propagation of equatorially trapped Kelvin and Rossby waves \citep{ShowmanPolvani11}, heat transport by the eastward zonal flow \citep[e.g.,][]{Cowan&Agol11,Zhang&Showman16}, and Doppler-shifted Kelvin and Rossby waves due to the eastward zonal flow \citep{Hammond&Pierrehumbert18}. 
In the context of our nonsynchronized, rapidly rotating planets, the displacement is caused by the time delay of the height field response to stellar irradiation, also seen in other literature for nonsynchronized planets \citep{Showman+15,Penn&Vallis17,Penn&Vallis18}.
When the substellar point moves westward due to a rapid rotation ($P_{\rm rot}<P_{\rm orb}$), the hot spot is shifted eastward from the substellar point unless the gravity wave is faster than the substellar point movement \citep{Penn&Vallis17}.
In our simulations, the gravity wave speed ($\sqrt{gH}{\sim}450~{\rm m~s^{-1}}$) is much slower than the substellar point velocity ($2\pi R_{\rm p}|P_{\rm rot}^{-1}-P_{\rm orb}^{-1}|{\sim}12000~{\rm m~s^{-1}}$).
Therefore, the height field always exhibits an eastward displacement of the hot spot from the substellar point when the radiative timescale is short in regime I.

For planets with $\theta>0^{\circ}$, the height fields exhibit seasonal variations, basically following the substellar point movement.
At the equinox where the substellar point is located at the equator, the shape of the height field is similar to that on nontilted planets.
At the solstice, the atmosphere undergoes an intense heating in the illuminated hemisphere if the obliquity is high.
As a result, the height field at the solstice is considerably different from that at the equinox.

The dynamical pattern for tilted planets is driven by a time-varying stellar insolation.
The flow pattern at the equinox is nearly independent of obliquity and looks similar to that on nontilted planets.
But the flow pattern at the solstice is different from that on nontilted planets when the obliquity is high.
For $\theta\leq{30}^{\circ}$, the flow pattern at the solstice shows a day-to-night flow pattern and is roughly similar to that on nontilted planets (the second row in Figure \ref{fig:hf_e0t05}).
However, in the cases of $\theta={60}^{\circ}$ and ${90}^{\circ}$, the flow patterns are dominated by westward winds on the illuminated hemispheres (the third and bottom rows in Figure \ref{fig:hf_e0t05}).
The westward flows are caused by the balance between the Coriolis force and the pressure gradient from the illuminated pole to the equator.

\subsubsection{Regimes II and III: Diurnal Mean Heating Is Dominant}\label{sec:regime23_e0}
The atmospheric dynamics are controlled by diurnally averaged insolation for $\tau_{\rm rad}=5~{\rm days}$, which is longer than the rotation period but shorter than the orbital period (Figure \ref{fig:hf_e0t5}).
We find that the height fields and flow patterns for $\theta={30}^{\circ}$, ${60}^{\circ}$, ${90}^{\circ}$ are considerably different from those for $\theta={0}^{\circ}$, ${10}^{\circ}$.
This is consistent with the obliquity criterion of $\theta\approx{18}^{\circ}$ predicted in Section \ref{sec:Regime} and explains the transition between the two dynamical regimes.

The atmospheric dynamics exhibit seasonal variations when planetary obliquity is larger than ${18}^{\circ}$, as expected in Section \ref{sec:Regime}.
For lower-obliquity cases ($\theta=0^{\circ}$ and ${10}^{\circ}$), the height fields and flow patterns are qualitatively similar for both cases (the first and second rows in Figure \ref{fig:hf_e0t5}). 
The height field is longitudinally homogenized and maximized at the equator, leading to eastward flows on the entire planet throughout the planet orbit.
On the other hand, for higher-obliquity cases with $\theta = {30}^{\rm \circ}$ and ${90}^{\rm \circ}$, the height field is highly time-dependent and higher in the illuminated polar region than at the equator (the third and bottom rows of Figure \ref{fig:hf_e0t5}).
This results in westward flows on the illuminated hemisphere and eastward flows on the other hemisphere in geostrophic balance.

Although our simulations are based on a simple shallow water system and Newtonian cooling scheme, our results are qualitatively consistent with the 3D simulations performed by \citet{Rauscher17}.
Especially, the two studies show a similar time evolution of the temperature structures (height fields in our cases) for both low and high obliquities.
In general, the 2D wind patterns from our simulations are also consistent with that in \citet{Rauscher17}, which exhibits steady eastward flows for lower-obliquity cases ($\theta={0}^{\rm \circ}$ and ${10}^{\rm \circ}$) and seasonally varying flow patterns for higher-obliquity cases ($\theta\geq{30}^{\circ}$), including the westward flows for $\theta={60}^{\rm \circ}$ \citep[see Figure 1 of][]{Rauscher17}.
But our simulations with $\theta={90}^{\circ}$ (bottom row in Figure \ref{fig:hf_e0t5}) are different from the 3D model results in \citet[][bottom row of Figure 1]{Rauscher17}.
We show alternating eastward and westward jetss, but \citet{Rauscher17} shows westward flows on the entire planet.
This is possibly related to the height field distributions in our simulations, which are also slightly different from that in \citet{Rauscher17}. 
For the $\theta={90}^{\circ}$ case, our simulations show that the winter pole is colder than the equator at the solstice, while the winter pole is slightly hotter than the equator in \citet{Rauscher17}. 
This discrepancy might be due to the difference in the atmospheric response to the heating between our 2D simulations and the 3D simulations in \citet{Rauscher17}. 
As noted in her paper, the response time of the temperature pattern is an order of magnitude longer than the radiative timescale, due to a complex radiative and dynamical response in her 3D model. 
In fact, the westward flows on the entire planet simulated in \citet{Rauscher17} resemble the flow patterns in our regime V with a longer radiative timescale, introduced in the next section.

\subsubsection{Regimes IV and V: Annual Mean Heating Is Dominant}\label{sec:regime45_e0}
Seasonal variations of the height fields and flow patterns are less obvious when the radiative timescale is longer than the orbital period.
Figure \ref{fig:hf_e0t100} shows the height fields and flow patterns for $\tau_{\rm rad}=100~{\rm days}$, which is longer than both the rotation and orbital periods.
The height fields are almost longitudinally homogenized and constant with time.
It implies that the long radiative timescale makes both the diurnal cycle and the seasonal cycle less important.
We find that the height fields and flow patterns are considerably different between the cases with $\theta\leq{30}^{\circ}$ and $\theta\geq{60}^{\circ}$.
This is again in agreement with our obliquity criterion of $\theta={54}^{\circ}$ predicted in Section \ref{sec:Regime}.

For the lower-obliquity cases ($\theta={0}^{\rm \circ}$, ${30}^{\rm \circ}$), the annual mean insolation is maximized at the equator, and the atmospheric dynamics are driven by the pressure gradient from the equator to the poles (the top and second rows in Figure \ref{fig:hf_e0t100}).
As a result of geostrophic balance, the flow patterns are dominated by eastward flows on the entire planet in our simple 2D fluid model, which is similar to that in regime II (Section \ref{sec:regime23_e0}).
In the cases of $\theta={60}^{\rm \circ}$ (third row in Figure \ref{fig:hf_e0t100}), the insolation is higher at both the north and south poles than at the equator. 
This results in flow patterns dominated by westward jets in the midlatitude region and a weak eastward jet at the equator.
As obliquity increases, the flow pattern is eventually dominated by westward flows on the entire planet, as seen in the case of $\theta={90}^{\rm \circ}$ (the bottom row in Figure \ref{fig:hf_e0t100}).

\subsection{Planets with Retrograde Rotation}\label{sec:dynamic_retro}
As mentioned in Section \ref{sec:Regime}, the height fields and flow patterns on planets with retrograde rotation (i.e., $\theta>{90}^{\circ}$) are qualitatively similar to those on planets with obliquity ${180}^{\circ}-\theta$.
Figure \ref{fig:retro_summary} shows the height and flow fields on planets with $\theta={120}^{\circ}$, ${150}^{\circ}$, and ${180}^{\circ}$ for various radiative timescales.
For example, in the case of $\tau_{\rm rad}=0.1~{\rm day}$ (the left column in Figure \ref{fig:retro_summary}), the atmosphere shows a strong day--night contrast in the height field and a day-to-night flow pattern, which are also seen on planets with $\theta\leq{90}^{\circ}$ (Figure \ref{fig:hf_e0t05}).
The height and flow patterns on planets with obliquity $\theta$ look very similar to that with ${180}^{\circ}-\theta$ (see Figure \ref{fig:hf_e0t05}).
This is seen in the cases of $\tau_{\rm rad}=5$ and $100~{\rm days}$ as well.

The reason for the above mirror similarity is that the substellar point movements are similar between the $\theta$ and ${180}^{\circ}-\theta$ cases in our simulations.
For instance, in the case of $\theta={180}^{\circ}$, the substellar point moves on the equatorial plane with an angular velocity $2\pi(P_{\rm rot}^{-1}+P_{\rm orb}^{-1})$, which is faster than the case of $\theta=0^{\circ}$ in which an angular velocity is $2\pi(P_{\rm rot}^{-1}-P_{\rm orb}^{-1})$.
However, since the rotation period is much shorter than the orbital period in our simulations, both substellar velocities are approximately $2\pi/P_{\rm rot}$.
Therefore, planets with obliquity $\theta$ and ${180}^{\circ}-\theta$ experience nearly the same insolation evolution, leading to similar temperature and dynamical patterns.

\subsection{Dynamical Patterns on ET Planets}\label{sec:dynamic_ET}
\begin{figure*}[t]
\centering
\includegraphics[clip, width=\hsize]{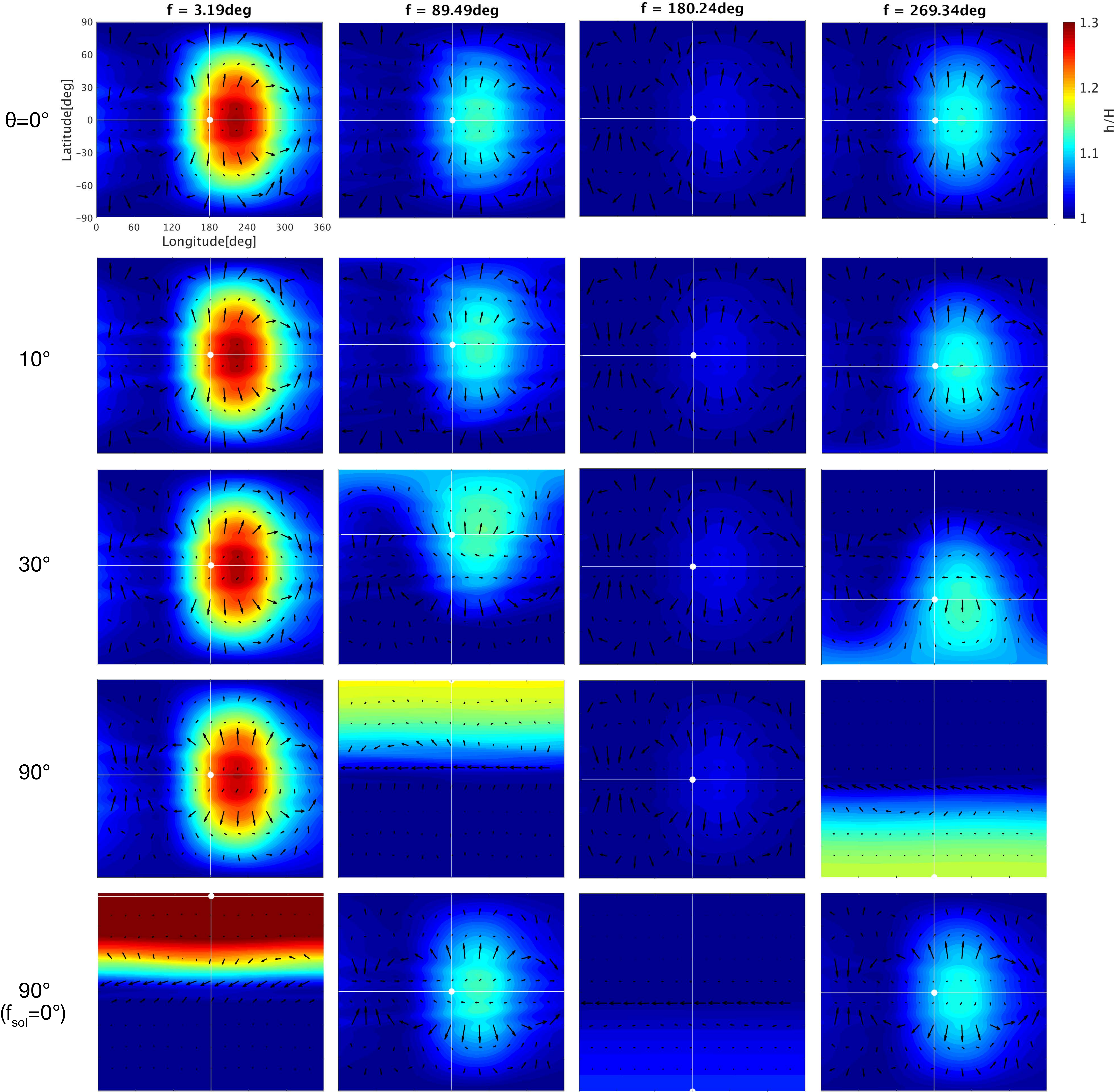}
\caption{Snapshots of height fields (color scale) and winds (arrows) for ET planets with $e=0.5$.
The radiative timescale is set to $\tau_{\rm rad}=0.1~{\rm day}$ in regime I. 
The horizontal axes are longitude with the substellar point (the white dot) at ${180}^{\circ}$.
From left to right, each column shows the snapshots taken near the true anomaly $f\approx 0^{\circ}$, ${90}^{\circ}$, ${180}^{\circ}$, and ${270}^{\circ}$, respectively.
From top to bottom, each row exhibits the snapshots for $\theta=0^{\circ}$, ${10}^{\circ}$, ${30}^{\circ}$, and ${90}^{\circ}$, respectively. 
Both of the last two bottom rows show the snapshots for $\theta={90}^{\circ}$, but the orbital phase of equinox is different: in the bottom row, the vernal equinox occurs a quarter year before the periapse passage.
Note that the lengths of arrows are normalized by the maximum length at each snapshot to clarify the dynamical patterns and do not represent magnitudes of the wind velocity. 
}
\label{fig:hf_e05t05}
\end{figure*}

\begin{figure*}[t]
\centering
\includegraphics[clip, width=\hsize]{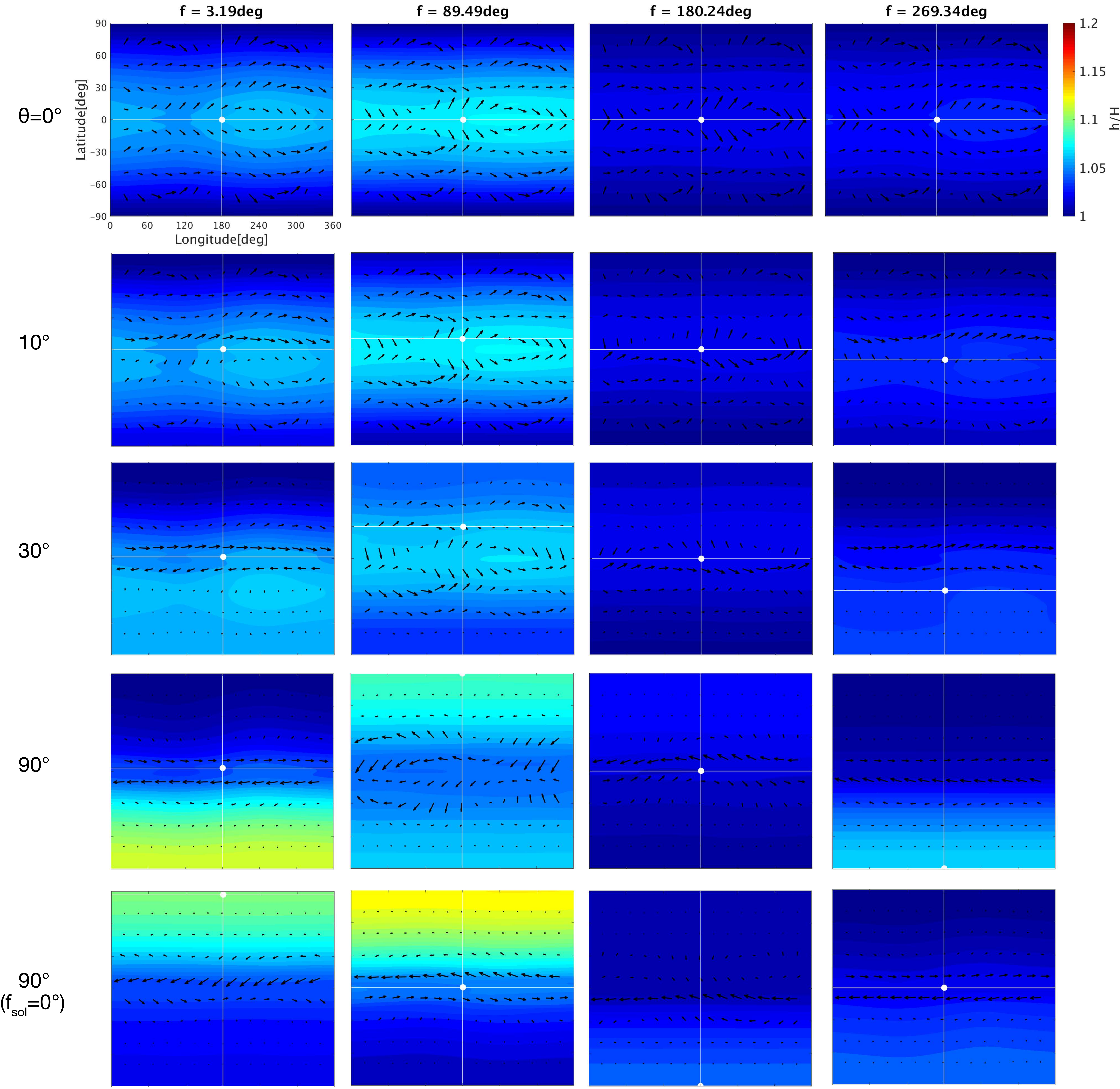}
\caption{Same as Figure \ref{fig:hf_e05t05}, but for $e=0.5$ and $\tau_{\rm rad}=5~{\rm days}$ in regimes II and III in which the diurnal mean heating is dominant.}
\label{fig:hf_e05t5}
\end{figure*}

\begin{figure*}[t]
\centering
\includegraphics[clip, width=\hsize]{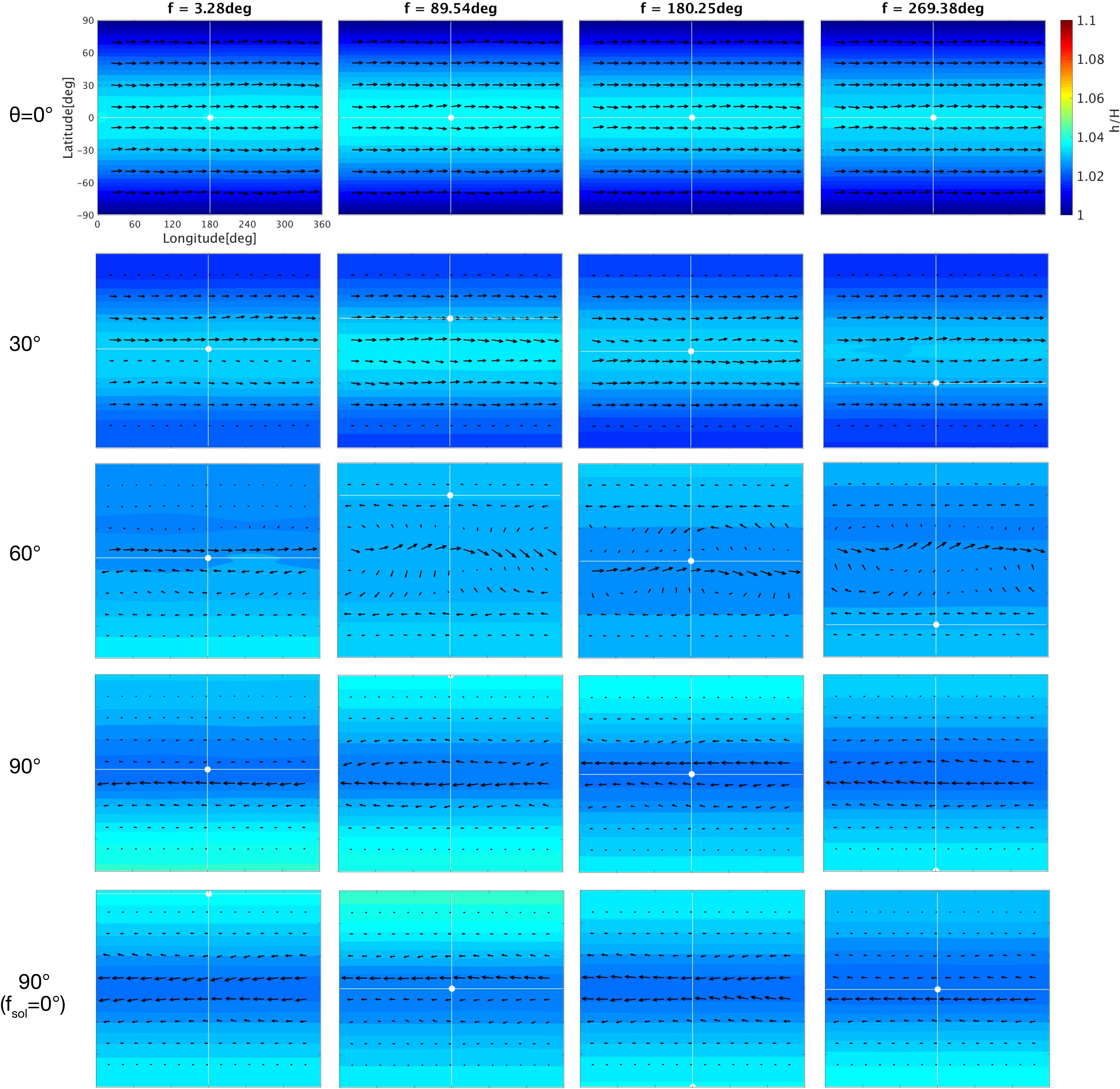}
\caption{Same as Figure \ref{fig:hf_e05t05}, but for $e=0.5$ and $\tau_{\rm rad}=100~{\rm days}$ in regimes IV and V in which the annual mean heating is dominant. From top to bottom, each row shows the snapshots for obliquity $0^{\circ}$, ${30}^{\circ}$, ${60}^{\circ}$, and ${90}^{\circ}$, respectively.}
\label{fig:hf_e05t100}
\end{figure*}

Now we move on to the dynamical patterns on ET planets.
Although we carried out simulations for $e=0.3$ and $e=0.5$, we found that dynamical patterns are qualitatively similar for these eccentricities, including the obliquity criteria for regime transition expected in Section \ref{sec:Regime}.
In addition, eccentricities are lower than $0.5$ for most detected exoplanets \citep[e.g.,][]{Winn&Fabrycky15}.
Therefore, we will only show the results for $e=0.5$ as representative cases for ET planets.
We show the dynamical patterns for a variety of obliquities and radiative timescales in Figures \ref{fig:hf_e05t05}--\ref{fig:hf_e05t100}.
The seasonal variation of the atmospheric dynamics depends on not only $e$ and $\theta$ but also on the northern summer solstice phase $f_{\rm sol}$, which makes the problem highly complicated. 
The most practical geometry might be $f_{\rm sol}=0^{\circ}$ and $f_{\rm sol}=\pm{90}^{\circ}$.
For $f_{\rm sol}=0^{\circ}$, in which the northern summer solstice takes place at the periapse, eccentricity operates on height fields to enhance the magnitude at the solstice.
On the other hand, for $f_{\rm sol}=0^{\circ}$, in which the vernal equinox takes place at the periapse, the eccentricity season competes with the obliquity season and might induce a seasonal variation of dynamical pattern distinct from that on circular-orbit planets.
Therefore, we will mainly show the results for $f_{\rm sol}={90}^{\circ}$ to focus on eccentricity effect and only show the results of $\theta={90}^{\circ}$ for $f_{\rm sol}={0}^{\circ}$.
As shown later, the regime classification presented in Section \ref{sec:Regime} is also applicable to the planets in eccentric orbits.
We discuss the results for different dynamical regimes in the following subsections. 

\subsubsection{Regime I: Day-Night Contrast Is Dominant}\label{sec:regime1_e05}
As noted, in an eccentric orbit, planets experience the ''eccentricity season'' in which the incoming stellar forcing changes as the star--planet distance varies.
This effect, which is independent of the obliquity effect, can be remarkably seen in the case of $\tau_{\rm rad}=0.1~{\rm day}$ (Figure \ref{fig:hf_e05t05}).
Because of the short radiative timescale in regime I, the atmosphere undergoes an intense heating near the periapse, and the wind at the periapse is stronger than that at the apoapse.
These behaviors are commonly seen in all obliquity cases in this regime.

The atmosphere exhibits a day-to-night flow pattern for $\theta=0^{\circ}$ and ${30}^{\circ}$ (the top and second rows in Figure \ref{fig:hf_e05t05}), and westward flow emerges on the illuminated hemisphere at the solstices for $\theta={60}^{\circ}$ and ${90}^{\circ}$ (the third and fourth rows in Figure \ref{fig:hf_e05t05}). 
These flow patterns are qualitatively similar to the results for circular-orbit planets (see Section \ref{sec:regime1_e0} and Figure \ref{fig:hf_e0t05}).
This is because the eccentricity only affects the magnitude, instead of the pattern, of the incoming stellar insolation.
Therefore, although the wind velocity varies with a star--planet distance, the flow pattern depends less on the eccentricity, as predicted in Section \ref{sec:Regime}.
Similar results are also seen in 3D simulations for eccentric hot Jupiters with pseudosynchronous rotation \citep{Kataria+13,Lewis+14}.

The seasonal variations of the height fields on planets in eccentric orbits depend not only on the obliquity but also on the phase angle between the periapse and the summer solstice, i.e., $f_{\rm sol}$.
The bottom row of Figure \ref{fig:hf_e05t05} shows the seasonal variations for planets with $\theta={90}^{\circ}$ where the northern summer solstice occurs at the periapse ($f_{\rm sol}=0^{\circ}$).
In this architecture, the height fields are maximized at the solstice.
By contrast, if the vernal equinox occurs at the periapse ($f_{\rm sol}={90}^{\circ}$), the height field is maximized at the equinox, but the atmosphere is also subsequently heated at the solstice (the fourth row in Figure \ref{fig:hf_e05t05}).
This indicates that, in regime I, the height field is largely controlled by the temporary insolation at the equinox.

\subsubsection{Regimes II and III: Diurnal Mean Heating Is Dominant}\label{sec:regime23_e05}
In regimes II and III, the atmospheric dynamics also undergo the seasonal variations caused by the eccentric orbit (Figure \ref{fig:hf_e05t5}).
The top row of Figure \ref{fig:hf_e05t5} shows the seasonal variations of the height fields and flow patterns for $\theta=0^{\circ}$ and $\tau_{\rm rad}=5~{\rm days}$. 
The remarkable difference from regime I is that the height fields are maximized considerably after the periapse passage, due to a relatively long radiative timescale.
The circulation is dominated by eastward flows on the entire planet, similar to the cases of circular-orbit planets (Section \ref{sec:regime23_e0}, Figure \ref{fig:hf_e0t5}).
The flow pattern near the equator fluctuates more because the irradiation magnitude changes with time in an eccentric orbit.

We find that the obliquity criterion $\theta={18}^{\circ}$ predicted in Section \ref{sec:Regime} also controls the dynamical patterns on ET planets.
Figure \ref{fig:hf_e05t5} shows that the flow pattern is dominated by eastward flows on the entire planet for $\theta={10}^{\circ}$, while it exhibits westward flows on the illuminated hemisphere and eastward flows on the other hemisphere for $\theta={30}^{\circ}$.
Again this is because the eccentricity only affects the magnitude of the insolation, but the spatial distribution of the diurnally averaged insolation is independent of eccentricity (see Equation \ref{eq:diurnal}).

The orbital phase difference between the vernal equinox and the periapse might have a significant influence on the circulation patterns on ET planets in this regime.
This is because the planet moves very rapidly around the periapse, so the orbital timescale becomes temporally comparable to the radiative timescale.
This will induce a temporal transition from regime III to V.
Integrating Kepler's equation (Equation \ref{eq:Kepler2}), one can evaluate the half-orbit timescale around the periapse $\tau_{\rm half}$, defined as the duration a planet takes to travel from $f=-\pi/2$ to $\pi/2$ (around the periapse): 
\begin{equation}\label{eq:P_half}
\tau_{\rm half}=\frac{P_{\rm orb}}{\pi}\left[ 2\tan^{-1}{\sqrt{\frac{1-e}{1+e}}}-e\sqrt{1-e^2}\right].
\end{equation}
The half-orbit timescale is $\tau_{\rm half}{\sim}0.2P_{\rm orb}$ for $e=0.5$ and $\tau_{\rm half}{\sim}0.02P_{\rm orb}$ for $e=0.9$.
For example, the fourth and bottom rows of Figure \ref{fig:hf_e05t5} show the seasonal variations of the height fields and flow patterns on planets with $\theta={90}^{\circ}$ in eccentric orbits with $f_{\rm sol}={90}^{\circ}$ and ${0}^{\circ}$, respectively.
For $f_{\rm sol}=0^{\circ}$, the shapes of the height fields and flow patterns are qualitatively similar to those on circular-orbit planets (Section \ref{sec:regime23_e0}, Figure \ref{fig:hf_e0t5}).
However, in the case of $f_{\rm sol}={90}^{\circ}$, the height fields on both poles become hotter than the equator at the northern summer solstice.
Westward flows emerge on the entire planet, similar to the flow patterns in regime V (Section \ref{sec:regime45_e0}).
This is because the dynamical pattern is temporarily dominated by the mean insolation around the periapse from the winter solstice all the way to the summer solstice, in which the planet travel timescale ($\tau_{\rm half}{\sim}6~{\rm days}$) is comparable to the radiative timescale ($\tau_{\rm rad}=5~{\rm days}$).

The other effect of the high eccentricity is that the radiative timescale experiences a significant temporal variation with star--orbit distance in a highly eccentric orbit, which could result in the distinct dynamical patterns.
\citet{Lewis+17} showed that the flow and temperature patterns on an extremely eccentric hot Jupiter HD 80606B ($e{\sim}0.93$) are significantly different between the periapse and the apoapse.
This is possibly due to different radiative timescales at different orbital phases.
The radiative timescale can be roughly estimated as \citep{ShowmanGuillot02}
\begin{equation}\label{eq:tau_rad}
\tau_{\rm rad}{\sim}\frac{P_{\rm ph}}{g}\frac{c_{\rm p}}{4\sigma T^3},
\end{equation}
where $P_{\rm ph}$ is the photospheric pressure, $T$ is the photospheric temperature, and $\sigma$ is the Stefan-Boltzmann constant.
Substituting the equilibrium temperature in Equation \eqref{eq:tau_rad}, we find the radiative timescale evolves with orbital phases as 
\begin{equation}\label{eq:tau_rad_e}
\tau_{\rm rad}{\sim}\frac{P_{\rm ph}}{g}\frac{c_{\rm p}}{4\sigma T_{\rm eq}^3}\left( \frac{1-e^2}{1+e\cos{f}}\right)^{3/2},
\end{equation}
where $T_{\rm eq}$ is the equilibrium temperature for a circular-orbit planet at the same semi-major axis. 
Note that Equation \eqref{eq:tau_rad_e} can only provide a crude estimate of the radiative timescale because it assumes atmospheric temperature is at the radiative equilibrium at each orbital phase, which is attained only in the limit of very short radiative timescales.
Equation \eqref{eq:tau_rad_e} implies that the radiative timescale changes from the periapse ($f=0$) to the apoapse ($f=\pi$) by a factor of up to $[(1+e)/(1-e)]^{3/2}$, which is ${\sim}5$ for $e=0.5$ and ${\sim}80$ for $e=0.9$.
Thus, for extremely eccentric planets (e.g., $e=0.9$), the radiative timescale varies with orbital phase by more than one order of magnitude, which might change the interaction among $\tau_{\rm rad}$, $P_{\rm rot}$, and $P_{\rm orb}$ at different orbital phases and thus the dynamical regimes in Figure \ref{fig:regime}.
However, for the parameter space examined in this study ($e\leq0.5$), the radiative timescale varies by less than an order of magnitude and thus is likely insufficient to cause the regime transition.
In the Appendix \ref{sec:appendix}, we also demonstrated that the temporal and spatial variation of $\tau_{\rm rad}$ has only minor effects on the dynamical patterns in our study.

\subsubsection{Regimes IV and V: Annual Mean Heating Is Dominant}
The time variation of the irradiation level in an eccentric orbit is less important for atmospheres with sufficiently long radiative timescales, as seen in the case of $\tau_{\rm rad}={100}~{\rm days}$ (Figure \ref{fig:hf_e05t100}).
The top row of Figure \ref{fig:hf_e05t100} shows that the height fields at different orbital phases for $\theta=0^{\circ}$ are nearly constant throughout the planet orbit. 
This behavior clearly indicates that the height fields are controlled by the annual mean insolation, and thus the seasonal effect is very weak.
Therefore, both height fields and flow patterns are qualitatively the same as on planets in circular orbits.

The obliquity criterion ${54}^{\circ}$ for the dynamical regime transition from IV to V is also applicable to ET planets with a long radiative timescale.
The height fields are higher at the equator than both poles for $\theta=0^{\circ}$ and ${30}^{\circ}$ (the top and second rows of Figure \ref{fig:hf_e05t100}).
By contrast, both the north and south poles are hotter than the equator for $\theta={60}^{\circ}$ and ${90}^{\circ}$ (the third and bottom rows of Figure \ref{fig:hf_e05t100}).
Similar to circular-orbit planets, eastward flows dominate the entire planet for $\theta < {54}^{\circ}$, while westward flows dominate for $\theta > {54}^{\circ}$.
In addition, the flow patterns are independent of $f_{\rm sol}$ as shown in the middle and bottom rows of Figure \ref{fig:hf_e05t100}.
To summarize, the impact of eccentricity on height fields and flow patterns is less important for planets in the regime controlled by the annual mean insolation.

\section{Summary and Discussion}\label{sec:summary}
We have investigated the atmospheric dynamics on ET planets using a one-and-half-layer shallow water model.
We found that the dynamical patterns can be classified into the five regimes (Section \ref{sec:Regime}) in terms of planetary obliquity $\theta$ and the radiative timescale $\tau_{\rm rad}$.
Our classification is applicable to both eccentric-orbit and circular-orbit planets.

(1) If the radiative timescale is shorter than the rotation period (regime I, Section \ref{sec:regime1_e0}), the height field is dominated by a time-varying day--night contrast, and the atmosphere produces a day-to-night flow pattern. 
The flow pattern at the equinox is similar to that for nontilted planets for all obliquities because the height field is mainly determined by the equilibrium field, which is independent of the obliquity at the equinox.
On the other hand, at the solstice, the flow pattern is considerably different from the equinox and is dominated by a westward flow on the illuminated hemisphere if the obliquity is substantially high, for example, $\theta={60}^{\circ}$ and ${90}^{\circ}$.

(2) When the radiative timescale is longer than the rotation period but shorter than the orbital period, the height field is dominated by the diurnal mean insolation (regimes II and III, Section \ref{sec:regime23_e0}). 
The transition obliquity between regime II and III is $\theta={18}^{\circ}$ (Section \ref{sec:Regime}).
For $\theta<{18}^{\circ}$ (regime II), the height field is longitudinally uniform and maximized at the equator throughout the planet orbit, leading to an eastward flow on the entire planet.
For $\theta>{18}^{\circ}$ (regime III), the height field is strongly heated in the polar region on the illuminated hemisphere, leading to a westward flow on the illuminated hemisphere and an eastward flow on the other hemisphere.

(3) The height field is dominated by the annual mean insolation if the radiative timescale is longer than the orbital period (regimes IV and V, Section \ref{sec:regime45_e0}).
The transition obliquity between regime IV and V is $\theta={54}^{\circ}$ (Section \ref{sec:Regime}).
For $\theta<{54}^{\circ}$ (regime IV), the height field is maximized at the equator, and eastward flow emerges on the entire planet, which is similar to that in regime II.
For $\theta>{54}^{\circ}$ (regimeV), the height field is maximized at both northern and southern poles throughout the planet orbit, leading to a westward flow on the entire planets.
Because of a very long radiative timescale, the flow and temperature patterns are nearly invariant throughout the planet orbit.

(4) The dynamical regime presented in Section \ref{sec:Regime} is also applicable to eccentric planets (Section \ref{sec:dynamic_ET}).
Although wind velocity varies with the star--planet distance, the flow patterns are qualitatively similar to that on circular-orbit planets in each regime.
This is because the eccentricity only affects the magnitude of incoming stellar flux, but the spatial distribution of the insolation is independent of eccentricity.
However, for extremely eccentric planets, the eccentricity might induce the transition from regime III to regime V due to rapid planet motion around the periapse. 

We adopted an idealized shallow water model that is useful for investigating a large parameter space but involves some caveats because of its simplicity.
For example, we have used a one-and-a-half-layer shallow water model missing the baroclinic dynamics that are responsible for fast-rotating planets with a large meridional temperature gradient \citep[e.g.,][]{Showman+15}.
Since a large obliquity naturally produces a large meridional temperature gradient, the dynamics driven by the baroclinic instability can be important for the jet formation and heat transport \citep{Vallis06}.
Vertical heat transport is also not taken into account in our model, which may be important for temperature evolution, as suggested by \citet{Rauscher17}.
Our future study with 3D general circulation model will investigate how the three-dimensional nature affects the circulations on ET planets.

We adopted the Newtonian cooling scheme with a spatially and temporally constant radiative timescale in each case, although the actual radiative timescale varies with time and space, as mentioned in Section \ref{sec:regime23_e05}.
To check the effects of the varying radiative timescale, we have performed several test simulations using radiative timescales scaled by the height field in the Appendix \ref{sec:appendix}.
In the parameter space examined in this study, the circulation patterns are less influenced by the varying radiative timescale.
However, models with realistic radiative transfer would be needed to investigate the circulation on an extremely eccentric planet (not focused on this study because of its rarity) where the temperature significantly varies with orbital phase, as suggested for hot Jupiter HD 80606 b with eccentricity $e=0.93$ \citep{Lewis+17}.

We have also neglected the mechanical wave forcing from deep convective layers, known as an important mechanism for atmospheric circulations on giant planets in our solar system and brown dwarfs \citep{Conrath+90,West+92,Lian&Showman10,Zhang&Showman14,Showman+18}.
As mentioned in Section \ref{sec:method}, the mechanical wave forcing likely has minor effects because the internal heat flux could be much weaker than the incoming external flux.
However, it might be important if planetary eccentricity is so high that the stellar flux becomes weaker than the internal flux at around the apoapse.
It would be worth investigating the effects on dynamical regimes for highly eccentric planets in the future.

We showed that the temperature patterns could be drastically influenced by the obliquity.
Because light curve observations probe the horizontal distributions of temperature, it may offer signatures of nonzero obliquity of exoplanets.
In a subsequent paper \citep[Paper II,][]{Ohno&Zhang19}, we will thoroughly investigate the thermal light curves from our shallow water simulations over a range of obliquity, eccentricity, radiative timescale, and viewing geometry.


\acknowledgments
We thank the anonymous referee for constructive comments and Gongjie Li and Daniel Fabrycky for helpful discussions.
This work was mainly carried out at the Kavli Summer Program in Astrophysics 2016.
We acknowledge Pascal Garaud, Jonathan Fortney, and the entire Kavli scientific organizing committee for thorough support.
This work was supported by the Kavli Foundation, the National Science Foundation, the Other Worlds Laboratory at UCSC, and the University of California Santa Cruz.
K.O. was supported by JSPS KAKENHI Grant Numbers JP15H02065, JP16K17661, JP18J14557, JP18H05438, and Foundation for Promotion of Astronomy.
X.Z. was supported by NASA Solar System Workings grant NNX16AG08G and NSF Solar and Planetary Research grant AST1740921. 
Most simulations of this study were carried out on the UCSC Hyades supercomputer.

\appendix

\section{Test of Varying Radiative Timescale}\label{sec:appendix}
\begin{figure*}[t]
\centering
\includegraphics[clip, width=\hsize]{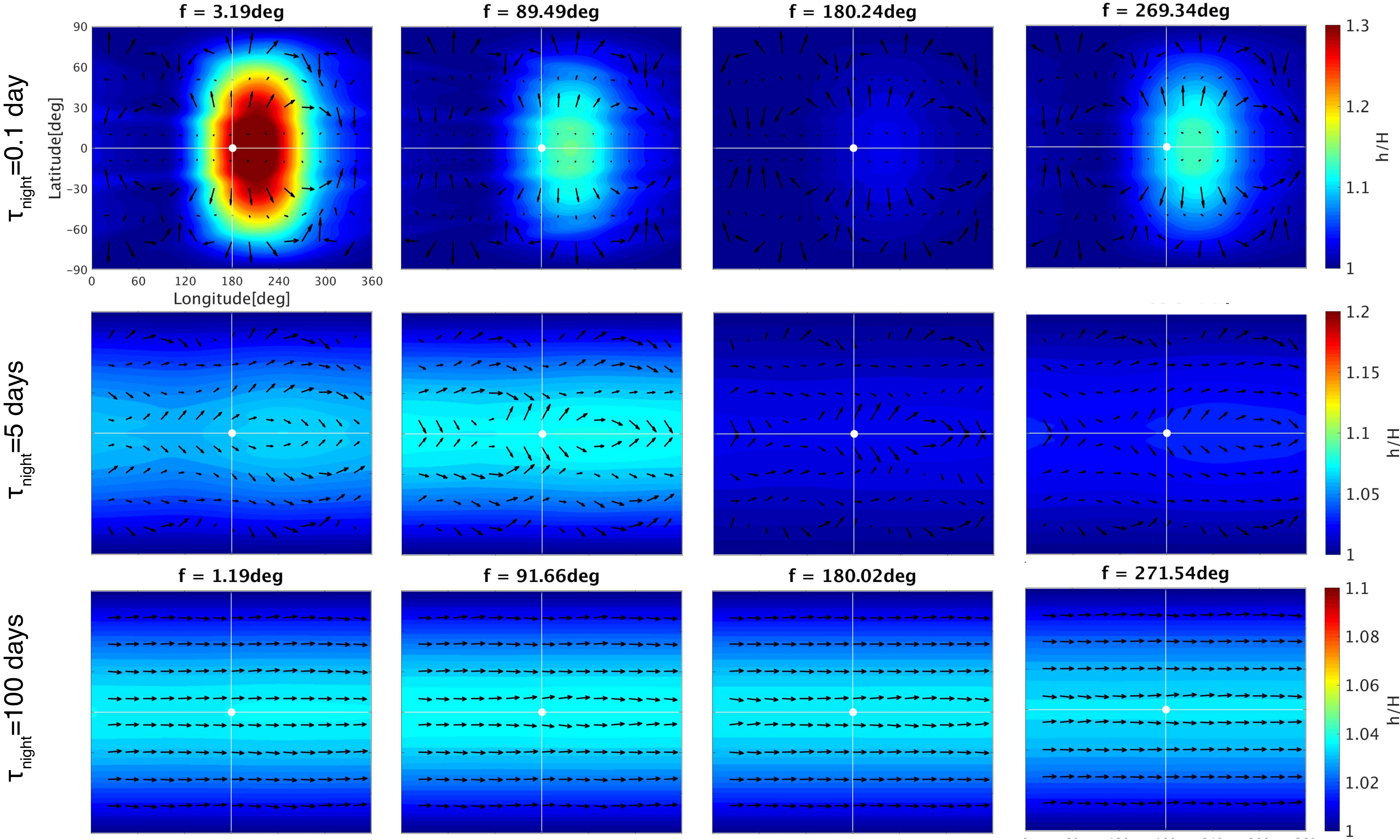}
\caption{Snapshots of height fields (color scale) and winds (arrows) for eccentric planets with zero obliquity from simulations with the prescribed radiative timescale (Equation \ref{eq:tau_rad_variable}). We set $\theta={0}^{\circ}$ and $\Delta h/H=0.1$. From left to right, each column shows the snapshots for true anomaly $f\approx 0^{\circ}$, ${90}^{\circ}$, ${180}^{\circ}$, and ${270}^{\circ}$, respectively. From top to bottom, each row shows the snapshots for a radiative timescale $\tau_{\rm night}=0.1$, $5$, and $100~{\rm days}$, respectively.
}
\label{fig:trad_test2}
\end{figure*}
To investigate the effects of temporally and spatially varying radiative timescales, we introduce the prescribed radiative timescale.
Assuming $T\propto h$, Equation \eqref{eq:tau_rad} might give the following scaling relation of the radiative timescale:
\begin{equation}\label{eq:tau_rad_variable}
\tau_{\rm rad}=\tau_{\rm night}\left( \frac{h}{H} \right)^{-3},
\end{equation}
where $\tau_{\rm night}$ is the mean radiative timescale on the nightside.
Although Equation \eqref{eq:tau_rad_variable} is crude, it allows us to examine how the dynamical patterns are influenced by temporally and spatially varying radiative timescales.

We perform several simulations for eccentric planets using Equation \eqref{eq:tau_rad_variable}.
Figure \ref{fig:trad_test2} shows the dynamical patterns for planets with $\theta=0^{\circ}$ and $e=0.5$.
Although the atmosphere is more strongly heated at the periapse for $\tau_{\rm night}=0.1~{\rm day}$ as compared to the case for constant radiative timescale of $\tau_{\rm rad}=0.1~{\rm day}$, the circulation patterns are very similar to those simulated with constant radiative timescales (top rows in Figures \ref{fig:hf_e05t05}--\ref{fig:hf_e05t100}). 
The results originate from the fact that a maximal variation of the height field is a factor of $1.3$ for $\tau_{\rm rad}=0.1~{\rm day}$ and $\la 1.1$ for $\tau_{\rm rad}=5~{\rm days}$ and $100~{\rm days}$, leading to variations of the radiative timescales by factor of up to $\approx2$ in Equation \eqref{eq:tau_rad_variable}.
The variation of a factor of $\approx2$ is inefficient to induce the transition of the dynamical regime.
3D simulations with sophisticated radiative transfer in \citet{Kataria+13} also showed that the time variation of globally averaged temperature is up to ${\sim}1.3$ (see Figure $9$ of their paper) and the flow pattern looks roughly invariable throughout the planet orbit.
Therefore, the dynamical pattern is likely less affected by the varying radiative timescale in the parameter space examined in this study.



\end{document}